\documentclass[reprint, superscriptaddress, aps, showpacs, prb]{revtex4-2}

\usepackage{graphicx}   
\usepackage{dcolumn}    
\usepackage{bm}         
\usepackage{amsmath,amsbsy,amsfonts,revsymb4-2}

\usepackage[justification=justified,format=plain]{caption}
\usepackage[textfont=normalfont,singlelinecheck=off]{subcaption}

\usepackage[utf8]{inputenc}

\DeclareCaptionJustification{justified}{\leftskip=0pt \rightskip=0pt \parfillskip=0pt plus 1fil}
\DeclareRobustCommand{\rchi}{{\mathpalette\irchi\relax}}
\newcommand{\irchi}[2]{\raisebox{\depth}{$#1\chi$}}

\makeatletter
\renewcommand{\p@subsection}{}
\renewcommand{\p@subsubsection}{}
\makeatother

\begin{document}

\title{Microscopic magnetism of $A$(TiO)Cu$_4$(PO$_4$)$_4$ ($A$ = Ba, Pb, Sr): $^{31}$P and $^{63,65}$Cu NMR study}

\author{Riho Rästa}
\email{riho.rasta@kbfi.ee}
\affiliation{National Institute of Chemical Physics and Biophysics, Akadeemia tee 23, 12618 Tallinn, Estonia}
\affiliation{Department of Cybernetics, Tallinn University of Technology, Ehitajate tee 5, Tallinn, 19086, Estonia}
\author{Ivo Heinmaa}
\affiliation{National Institute of Chemical Physics and Biophysics, Akadeemia tee 23, 12618 Tallinn, Estonia}
\author{Joosep Link}
\affiliation{National Institute of Chemical Physics and Biophysics, Akadeemia tee 23, 12618 Tallinn, Estonia}
\affiliation{Department of Cybernetics, Tallinn University of Technology, Ehitajate tee 5, Tallinn, 19086, Estonia}
\author{Yusuke Kousaka}
\affiliation{Department of Physics and Electronics, Osaka Metropolitan University, Osaka, 599-8531, Japan}
\author{Tsuyoshi Kimura}
\affiliation{Department of Applied Physics, The University of Tokyo, Tokyo, 113-8656, Japan}
\author{Yoshihiko Ihara}
\affiliation{Department of Physics, Faculty of Science, Hokkaido University, Sapporo 060-0810, Japan}
\author{Kenta Kimura}
\affiliation{Department of Materials Science, Osaka Metropolitan University, Osaka, 599-8531, Japan}
\author{Raivo Stern}
\email{raivo.stern@kbfi.ee}
\affiliation{National Institute of Chemical Physics and Biophysics, Akadeemia tee 23, 12618 Tallinn, Estonia}
\affiliation{National High Magnetic Field Laboratory, 1800 E Paul Dirac Dr, Tallahassee, 32310, FL, USA}

\date{\today }
\begin{abstract}
	We report a comprehensive NMR study of the chiral square-cupola antiferromagnet Pb(TiO)Cu$_4$(PO$_4$)$_4$ and compare its microscopic hyperfine and local-field parameters with the Ba/Sr analogues in the $A$(TiO)Cu$_4$(PO$_4$)$_4$ family. Above $T_{\rm N}\simeq 6.7$~K, the $^{31}$P Knight shift tracks the bulk susceptibility and yields nearly isotropic transferred hyperfine couplings $H_{\rm hf}^{[010]}=6.77(3)$ and $H_{\rm hf}^{[001]}=6.19(3)$~kOe/$\mu_{\rm B}$. Below $T_{\rm N}$, the frequency-swept $^{31}$P spectrum splits into three lines, in contrast to the four-line pattern reported for BaTCPO. The line separation tracks the onset of the static $^{31}$P internal field with a power-law exponent $\beta\simeq 0.23$, consistent with quasi-two-dimensional criticality. Crystal-rotation $^{31}$P NMR in the ordered state resolves all eight symmetry-related P sites and their site-dependent anisotropy. In the ordered state, zero-field $^{63,65}$Cu NMR gives a Cu-site internal field $B_{\rm int}=14.50(6)$~T and a quadrupole frequency $\nu_Q=32.72(5)$~MHz, while point-charge electric-field-gradient calculations including Sternheimer corrections yield an on-site Cu hole occupancy $n_d=0.20(4)$, consistent with a ligand-hole--dominated charge-transfer character. Comparing PbTCPO with BaTCPO and SrTCPO, we find that the transferred hyperfine coupling $H_{\rm hf}$ varies across the series, reflecting changes in local Cu--O--P covalency, whereas the ordered-state $^{31}$P internal field in PbTCPO is $69.5$~mT, considerably higher than in BaTCPO ($35.6$~mT) and SrTCPO ($34.6$~mT). This enhancement is not captured by dipolar terms alone and points to the combined effects of transferred contributions and stacking-dependent cancellation.
\end{abstract}

\maketitle
\section{Introduction}

Multiferroic and magnetoelectric materials, where magnetic and electric orders coexist and interact, are central to condensed-matter research because of their potential for multifunctional spintronic and magneto-optical applications~\cite{Astrov1960,Dubovik1990,Kimura2003,Fiebig2005,cheong2007multiferroics,arima2008magneto,Spaldin2008,SpaldinNatMat18,mundy2016atomically,baltz2018antiferromagnetic}. Among the known classes of magnetoelectrics, antiferromagnets showing a linear magnetoelectric effect such as Cr${_2}$O${_3}$ have long been studied. In such systems, the macroscopic response is governed not by net magnetization, but by the symmetry-selected
spin configurations, conveniently discussed in terms of parity-odd magnetic multipoles and related cluster multipoles~\cite{Dubovik1990,Spaldin2008}.

The tetragonal square-cupola cuprates $A$(TiO)Cu$_4$(PO$_4$)$_4$ ($A$ = Ba, Sr, Pb), hereafter $A$TCPO, comprise layers of corner-sharing Cu$_4$O$_{12}$ square-cupola units linked by nonmagnetic Ti--O polyhedra~\cite{Kimura_inorganic}. The buckled geometry stabilizes a noncollinear antiferromagnetic spin arrangement on each cupola that is naturally expressed in terms of cluster multipoles, with a magnetic quadrupole as the relevant magnetoelectric component~\cite{Kato2019_SrTCPO,Kimura2018Acation}. Although the in-layer quadrupole pattern is ferroic, the $c$-axis stacking depends on the choice of \textit{A}: Ba/Sr realize antiferroic stacking due to antiferromagnetic interlayer couplings, whereas Pb stabilizes ferroic stacking via ferromagnetic interlayer couplings~\cite{Kato2019_SrTCPO,Kimura2018Acation}. Consequently, in PbTCPO, both time-reversal and space-inversion symmetries are broken globally, leading to a linear magnetoelectric effect and nonreciprocal linear dichroism, and enabling direct macroscopic probes and domain-sensitive optical techniques that are strongly suppressed in Ba/Sr~\cite{Kimura2016_magneto,Kimura2020}. In PbTCPO, chirality additionally induces a time-reversal-broken magnetic octupole component, which allows the control of magnetoelectric domains of the system, domain selection, and magnetic-field-only control protocols~\cite{Kimura2021,Kimura2022}. PbTCPO also hosts field-induced phases with distinct magnetoelectric responses~\cite{Kimura2019,Kimura2020}.

NMR provides a direct route to those novel phenomena on the microscopic level. In particular, stacking primarily controls interlayer cancellation in the $A$-site vector sum of dipolar and transferred-hyperfine fields, strongly affecting the ordered-state $^{31}$P internal fields, whereas the Cu--O--P bonding geometry sets the magnitude and anisotropy of the $^{31}$P hyperfine tensor. By contrast, the $^{63,65}$Cu zero-field internal field is dominated by the on-site hyperfine term. Establishing these local quantities is therefore essential for benchmarking microscopic models and for separating stacking-driven cancellation effects from local electronic-structure trends across the Ba/Sr/Pb series. Despite extensive macroscopic characterization~\cite{Kimura2016_magneto,Kimura2018Acation,Kimura2019,Kimura2020,Kimura2021,Nomura2023}, systematic microscopic study of all relevant parameters remains limited. While Ref.~\cite{Ihara2025a} focused primarily on the high-field $^{31}$P NMR phenomenology and field-induced states of PbTCPO, the present work focuses on the low-field and zero-field microscopic parameters and their comparison across the ATCPO family. Here, we use single-crystal $^{31}$P NMR, together with zero-field $^{63,65}$Cu NMR, to determine site-resolved hyperfine tensors, ordered-state internal fields, and Cu-site EFG parameters in PbTCPO, and compare them with established trends across $A$TCPO.

\section{Experiment}
Two single crystals of PbTCPO were used in this study. The first crystal (labelled \#1) sized $1.91\times2\times3~\text{mm}^3$ and weighed $31.51~\text{mg}$. A second sample (\#2), sized $0.5\times3\times4~\text{mm}^3$ and weighed $40.74~\text{mg}$, was grown later~\cite{Kimura2021} with its chirality confirmed by polarized-light microscopy \cite{Kimura2020} to be levorotatory. Unless stated otherwise, all PbTCPO single-crystal measurements reported below were performed on sample \#1. Magnetic susceptibility was measured using the vibrating-sample magnetometer (VSM) of the Quantum Design 14T-PPMS. $^{31}$P NMR measurements were conducted using the commercial Bruker AVANCE~II spectrometer with the sample placed in a $B_\text{ext}=4.7~\text{T}$ superconducting magnet (resonance frequency 80.97~\text{MHz}). A He-flow cryostat (Janis Research) enabled measurements over the temperature range 4.2--300~\text{K}. The NMR probe was home-built and equipped with a single-axis goniometer. Temperature was regulated using calibrated Cernox sensors and a Lake Shore Model~332 temperature controller. Spin--lattice relaxation $T_1$ was measured using the inversion-recovery sequence. The spectra were obtained using a frequency-sweep method. The Knight shifts are reported relative to the resonance of H$_3$PO$_4$. $^{63,65}$Cu NMR on the powder sample was performed in zero applied field using a liquid-He bath. All measurements were conducted at NICPB, Tallinn, Estonia. The BaTCPO values reported in this work were obtained from an extended reanalysis of the experimental data originally measured in Ref.~\cite{Rasta2020}. Programming tools used for calculations include MATLAB, C\texttt{++}, and Python. Unless noted otherwise, values in parentheses indicate the uncertainty in the last digit(s).

\section{Results and analysis}
\subsection{Magnetization}
\begin{figure}
	\centering
	\includegraphics[width=0.49\textwidth]{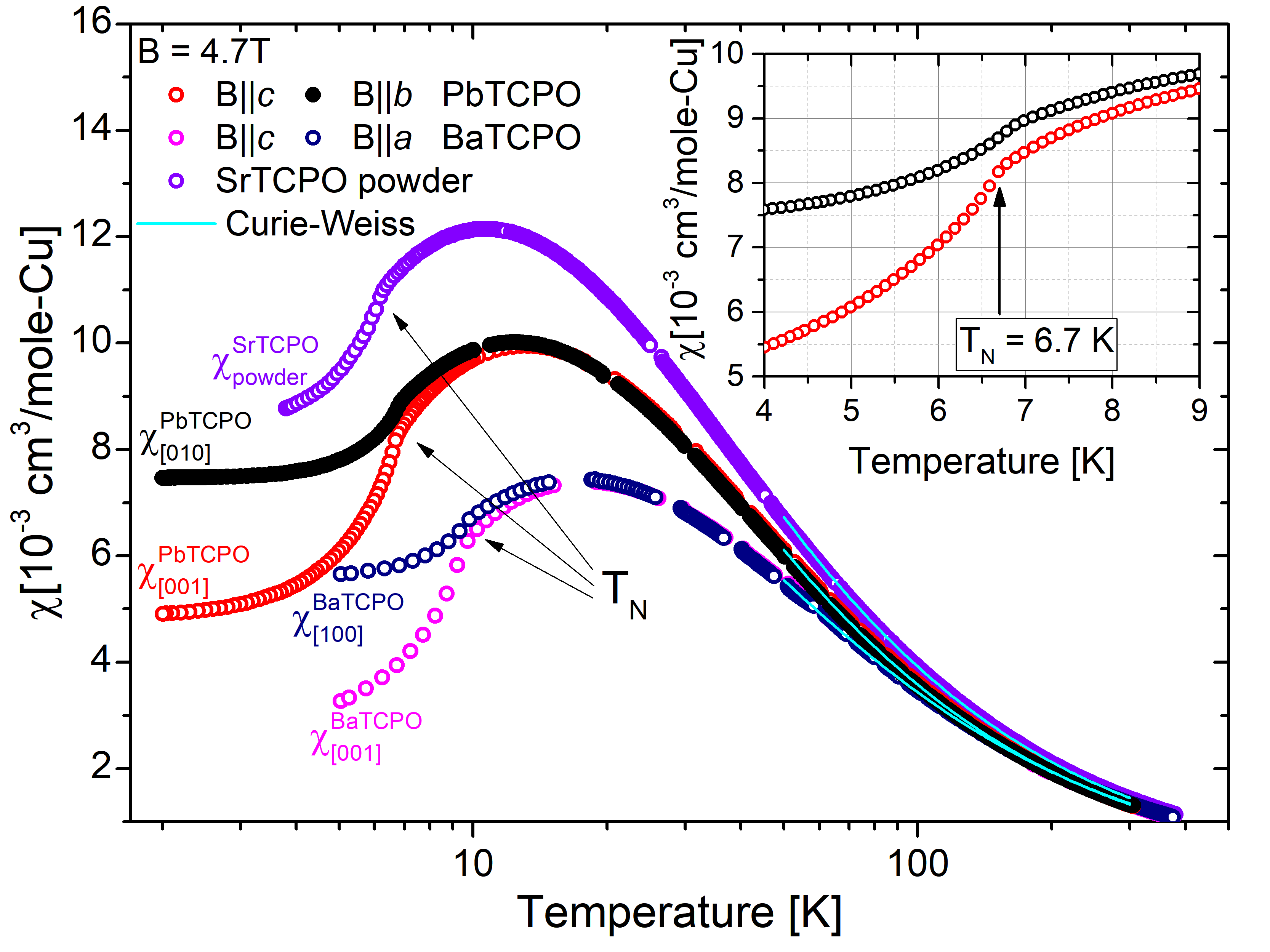}
	\caption{Temperature dependence of magnetic susceptibility $\rchi(T)$ of PbTCPO (black $B$$\parallel$[010] and red $B$$\parallel$[001]), BaTCPO (navy $B$$\parallel$[100] and magenta $B$$\parallel$[001]) single crystals, and SrTCPO (violet) powder sample in an applied magnetic field $B_{\mathrm{ext}}=4.7~\text{T}$. The cyan and green lines represent the Curie--Weiss fits for 50~-~300~K. The inset shows splitting at $T_{N}=6.7~\text{K}$ for PbTCPO.}
	\label{fig:susc}
\end{figure}

The temperature dependence of the magnetic susceptibility was measured at an external magnetic field $B_{\mathrm{ext}}=4.7~\text{T}$ for all three compounds (Fig.~\ref{fig:susc}). For $T>50~\text{K} $, $ \rchi(T)$ follows the Curie--Weiss relation:
\begin{align}
\rchi(T)=\rchi_0+\frac{C}{T-\theta_\text{CW}},
\end{align}
where $\rchi_0$ is the temperature-independent susceptibility and the second term is the Curie--Weiss law. The temperature-independent molar susceptibility per Cu was calculated by combining core diamagnetism $\rchi_D$, using Pascal's constants from Refs.~\onlinecite{Selwood1956, Bain2008}, and the Van Vleck paramagnetism $\rchi_\text{VV}$, which is approximately $5.0(12)\cdot 10^{-5}$ cm$^3$/mol \cite{Bleaney1952, Spodine2003, Barwiolek2020}, we get $\rchi_{0} = (\rchi_D + \rchi_\text{VV})/4 = -2.8(12) \cdot 10^{-5}$ cm$^3$/mol-Cu. Note that the parameters are sensitive to the selection of the temperature region (we chose $50\le T \le 300~\text{K}$ for all compounds). The fits can be seen on Fig.~\ref{fig:susc}; the fitting results are presented in Table \ref{tab:susc}.

The Curie constants are isotropic, 
while the Weiss parameters $\theta_{\text{CW}}$ show the strength of antiferromagnetic interactions between Cu$ ^{2+} $ ions. The small orientation dependence of \textit{C} indicates a nearly isotropic paramagnetic response. From the Curie constant we get the effective copper magnetic moment as  $\mu_{\text{eff}}=\sqrt{3k_BC/N_A}$, where $k_B$ is the Boltzmann factor, and $N_A$ is Avogadro's number, as well as an estimation for the Landé $g$-factor values from $g=\sqrt{S(S+1)}\mu_{\text{eff}}$. $g\approx2.16$--$2.25$ is consistent with Cu$^{2+}$ in a square-planar ligand field where partial orbital contribution and covalency enhance $g$ above 2 \cite{AbragamBleaney1970}. 
\begin{table}
	\vskip 15pt
	\caption{\label{tab:susc}Magnetic properties of $A$TCPO acquired from PPMS and $ ^{31} $P NMR measurements. The NMR results of SrTCPO are from Ref.~\onlinecite{Islam2018}.}
	\renewcommand{\arraystretch}{1.5}
	\resizebox{\columnwidth}{!}{%
	\begin{tabular}{|c|c|c|c|c|c|}
		\cline{2-6}
		\multicolumn{1}{c|}{ }& \multicolumn{3}{c|}{ PPMS } &\multicolumn{2}{c|}{$^{31}$P NMR} \\ \hline
		Compound & $C$          	& $\theta_\text{CW}(\text{K})$  & $g$ & $T_{N}$(K) & $H_\text{hf}$ $\mathrm{kOe}/\mu_{\mathrm{B}}$ \\ \hline
		BaTCPO [001] &  0.466(2) & -32.0(5) 	& 2.23(6) & 8.8 & 7.40(5)\\ \hline
		BaTCPO [100] &  0.464(1) & -32.8(2) 	& 2.22(6) & 9.5 & 7.65(5)\\ \hline
		PbTCPO [001] &  0.444(1) & -21.8(2)     & 2.17(5) & 6.7 & 6.19(3)\\ \hline
		PbTCPO [010] &  0.437(3) & -22.1(1) 	& 2.16(5) & 6.8 & 6.77(3)\\ \hline
		SrTCPO powder & 0.475(1) & -20.3(1) 	& 2.25(6) & 6.2 & 6.54$^\mathrm{a}$  \\ \hline
	\end{tabular}
    }
    \footnotetext[1]{From Ref.~\cite{Islam2018}.}
	\renewcommand{\arraystretch}{1}
\end{table}

In PbTCPO, $\rchi(T)$ stays almost identical for $B$$\parallel$[010] and $B$$\parallel$[001] down to $T_{N}$~=~6.7~K, where the curves split and $\rchi$ for $B$$\parallel$[001] decreases more rapidly. This is an indication of the magnetic spins of Cu$^{2+}$ ions turning perpendicular to the [010] direction and parallel to [001], similar to that in Ref.~\onlinecite{Singer1956}.

\begin{figure}
\centering
\includegraphics[width=0.5\textwidth]{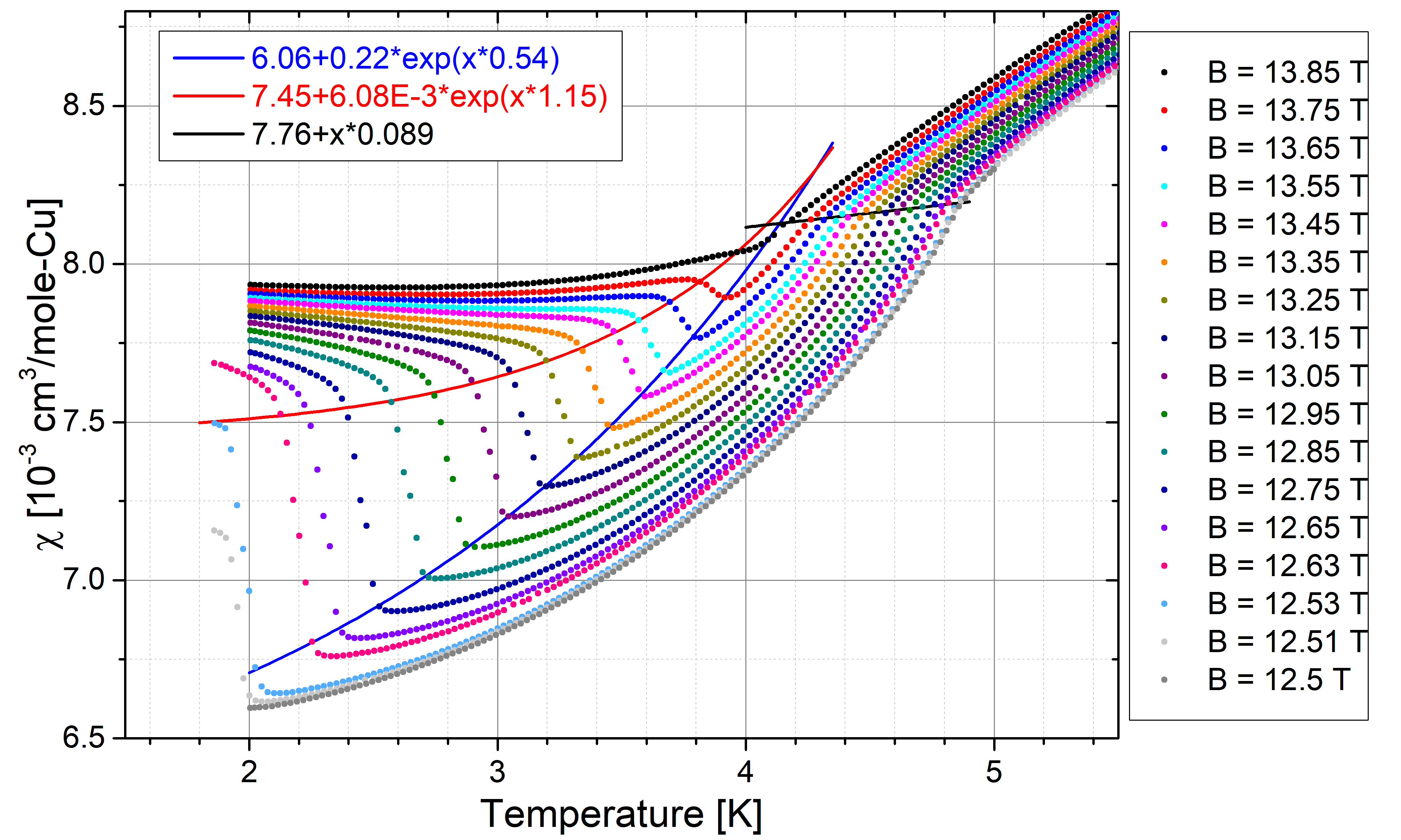}
\caption{PbTCPO temperature dependence of magnetic susceptibility $\rchi(T)$ at the field-induced phase transition region at varying magnetic fields, measured in the direction $B$$\parallel$[001].}
\label{fig:FI}
\end{figure}

\begin{figure}
	\centering
	\includegraphics[width=0.5\textwidth]{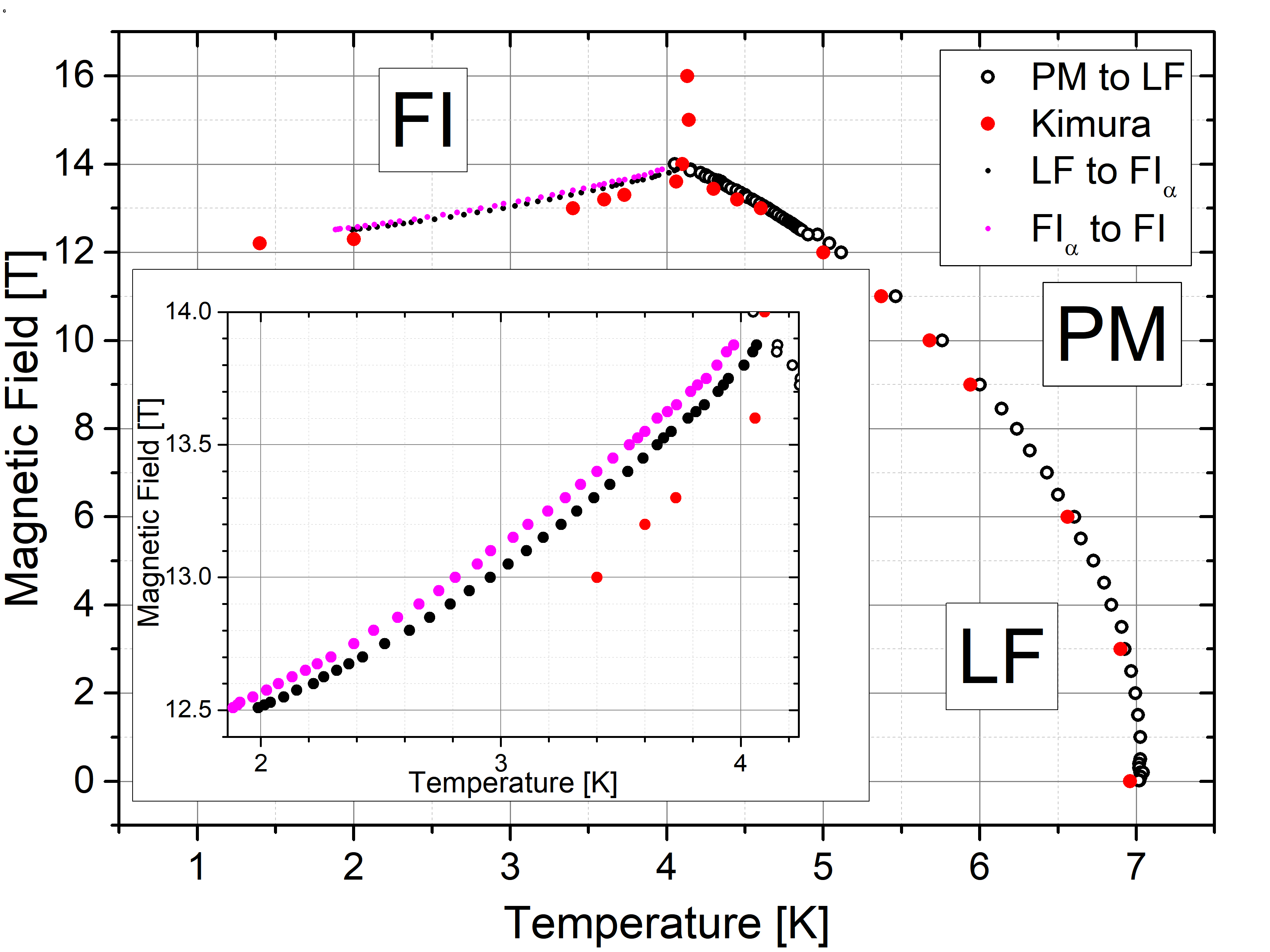}
	\caption{Magnetic field dependence of the phase transition temperature in PbTCPO with the single crystal $c$ axis ([001]) aligned with $B$. The black circles show the phase transition $T_N$ from room temperature to the ordered state; the black dots and pink dots show a two-step transition from the low-field (LF) ordered state to the field-induced (FI) ordered state. The red dots are from Ref.~\onlinecite{Kimura2019} for comparison.}
	\label{fig:phase}
\end{figure}

$\rchi(T)$ in PbTCPO with $B$$\parallel$[001] was measured in a range of external magnetic fields $B_\text{ext}$ from 0 to 14~\text{T}. A phase diagram is shown in Fig.~\ref{fig:phase}. The field-induced (FI) transition region was reported between 12.2--13.6~\text{T} in Ref.~\onlinecite{Kimura2019} while the FI phase transitions in our results start near 12.5~\text{T} at $T$ = 2~K, and evolve up to 13.9~\text{T}, where three transition lines merge. A close-up of the $\rchi(T)$ curves at the FI phase transition region is shown in Fig.~\ref{fig:FI}. For $B$ up to 14~T, the first phase transition is from the paramagnetic (PM) state to the low-field (LF) ordered state; at lower temperatures, a transition to the field-induced (FI) ordered state occurs. By decreasing the temperature, the LF to FI phase transition is characterized by an increase in $\rchi$, as shown by the onset (black dots in Fig.~\ref{fig:phase}) and the end (magenta dots in Fig.~\ref{fig:phase}) of the phase transition. The exact temperatures of these points were determined from the derivatives of the measurement lines. The $\sim$2\% upward shift compared to \cite{Kimura2019} of the FI window is attributable to temperature or field calibration offset, hysteresis, and exact transition point choice criteria.

\subsection{$^{31}$P Knight shift}

\begin{figure}[ht!]
	\begin{subfigure}[b]{0.4\textwidth}
		\includegraphics[width=1\textwidth]{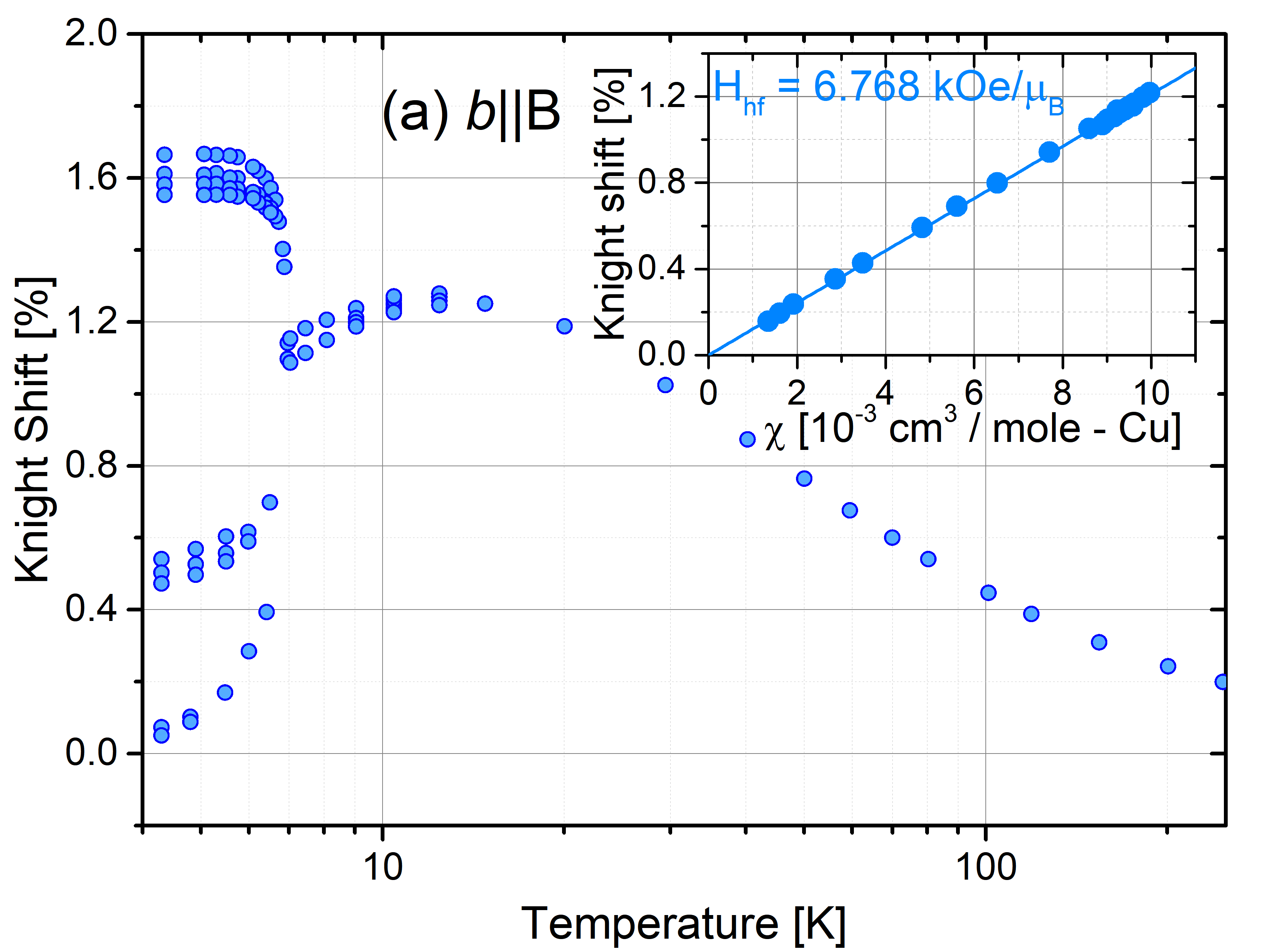}
	\end{subfigure}%
	
	\begin{subfigure}[b]{0.4\textwidth}
		\includegraphics[width=1\textwidth]{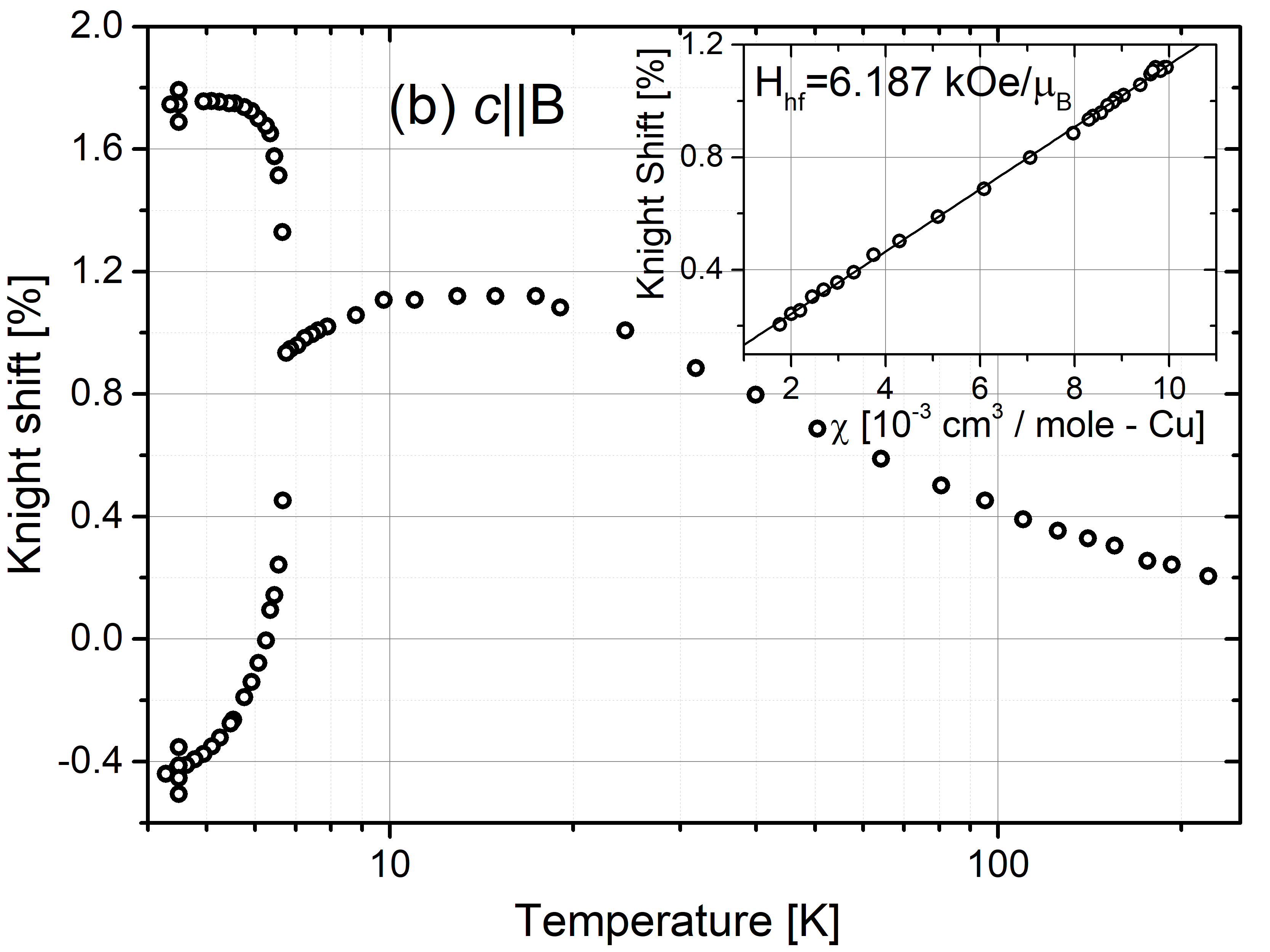}
	\end{subfigure}
	\caption{$^{31}$P Knight shift temperature dependence of PbTCPO single crystal for (a) $B$$\parallel$[010] and (b) $B$$\parallel$[001]. Insets: Clogston--Jaccarino plots $K ~\text{vs}~\rchi$ with linear fits; the slope gives $H_\text{hf}$.}
	\label{fig:Knight}
\end{figure}

The $^{31}$P Knight shift of PbTCPO was measured from room temperature down to 4.2~K (Fig.~\ref{fig:Knight}) with $B_{\rm ext}=4.7$~T applied along the $b$ axis ([010]) and along the $c$ axis ([001]). Above the Néel temperature $T_{N}$ = 6.7~K the Knight shift $K$ follows the magnetic susceptibility $\rchi$, as given by

\begin{equation} \label{C-J} 
K(T)=K_0+\frac{H_{\rm hf}}{N_A\mu_B}\ \rchi.
\end{equation}
\noindent
Here $K_0$ is the temperature-independent Knight shift, $\mu_B$ Bohr magneton, $N_A$ the Avogadro number, $H_{\rm hf}$ the hyperfine coupling and $\rchi$ is the molar susceptibility per Cu$^{2+}$. The Knight shift shows a broad maximum at $T_K^\text{max}\approx$~17~K, as does the magnetic susceptibility. The shift is larger by $\approx$0.2\% compared to BaTCPO in both orientations. From the $K$--$\rchi$ plots (Fig.~\ref{fig:Knight}a and \ref{fig:Knight}b insets) and Eq.~(\ref{C-J}) we obtained the hyperfine field values $H_{\rm hf}^{B\parallel b}$ = 6.77(3) $\mathrm{kOe}/\mu_{\mathrm{B}}$ and $H_{\rm hf}^{B\parallel c}$ = 6.19(4) $\mathrm{kOe}/\mu_{\mathrm{B}}$. These results are in good agreement with Ref.~\cite{Ihara2025a}. The results for all three compounds are summarized in Table \ref{tab:susc}. The near isotropy of $H_{\rm hf}$ indicates dominant Fermi-contact transfer at P, and the $A$-site dependence of $H_{\rm hf}$ reflects changes in Cu--O--P hybridization.

Below temperatures $T \le $ 15~K, the Knight shift anisotropy starts broadening the line and different $^{31}$P sites become distinguishable already in the paramagnetic state. At $T$ = 9~K, the lines split into two similar peaks. Between $T$ = 8~K and $T$ = 7.5~K, the higher-frequency peak is approximately half the intensity of the lower-frequency peak. At $T$ = 7~K, the lower intensity peak also decreases to the same lower intensity as the higher-frequency peak. 

At temperatures below $T_{N}$, the Knight shift resonance lines of PbTCPO single crystal split into two different projections for ${B\parallel c}$, higher-frequency $\nu_1$ and lower-frequency $\nu_2$, and into three projections for ${B\parallel b}$. Three projections were also measured by Ihara \textit{et al.} \cite{Ihara2025a}. This differs from BaTCPO, where the lines split into four with ${B\parallel a}$. The splitting $\Delta F$~=~$\nu_1 - \nu_2$, reflects the internal field at the P site from the ordered Cu$^{2+}$ moments ($\Delta F \propto \gamma_n B_{\mathrm{int}}$), and thus tracks the antiferromagnetic order parameter. Hence we fit

\begin{equation}\label{power}
\Delta F(T)=\Delta F_0\left(1-\frac{T}{T_{N}}\right)^\beta.
\end{equation}

The best fit was obtained with $T^c_{N} = 6.71~\text{K}$, $\beta\simeq0.22$ for ${B\parallel c}$ and $T_{N}^b = 6.77~\text{K}$, $\beta\simeq0.25$ for ${B\parallel b}$. Similarly to BaTCPO, the Néel temperature here is slightly lower with ${B\parallel c}$, but much more isotropic. 

The exponent $\beta$ is consistent with pronounced quasi-two-dimensional critical behavior \cite{Nath2009}.

\begin{figure}[t]
	\begin{subfigure}[b]{0.4\textwidth}
	\includegraphics[width=1\textwidth]{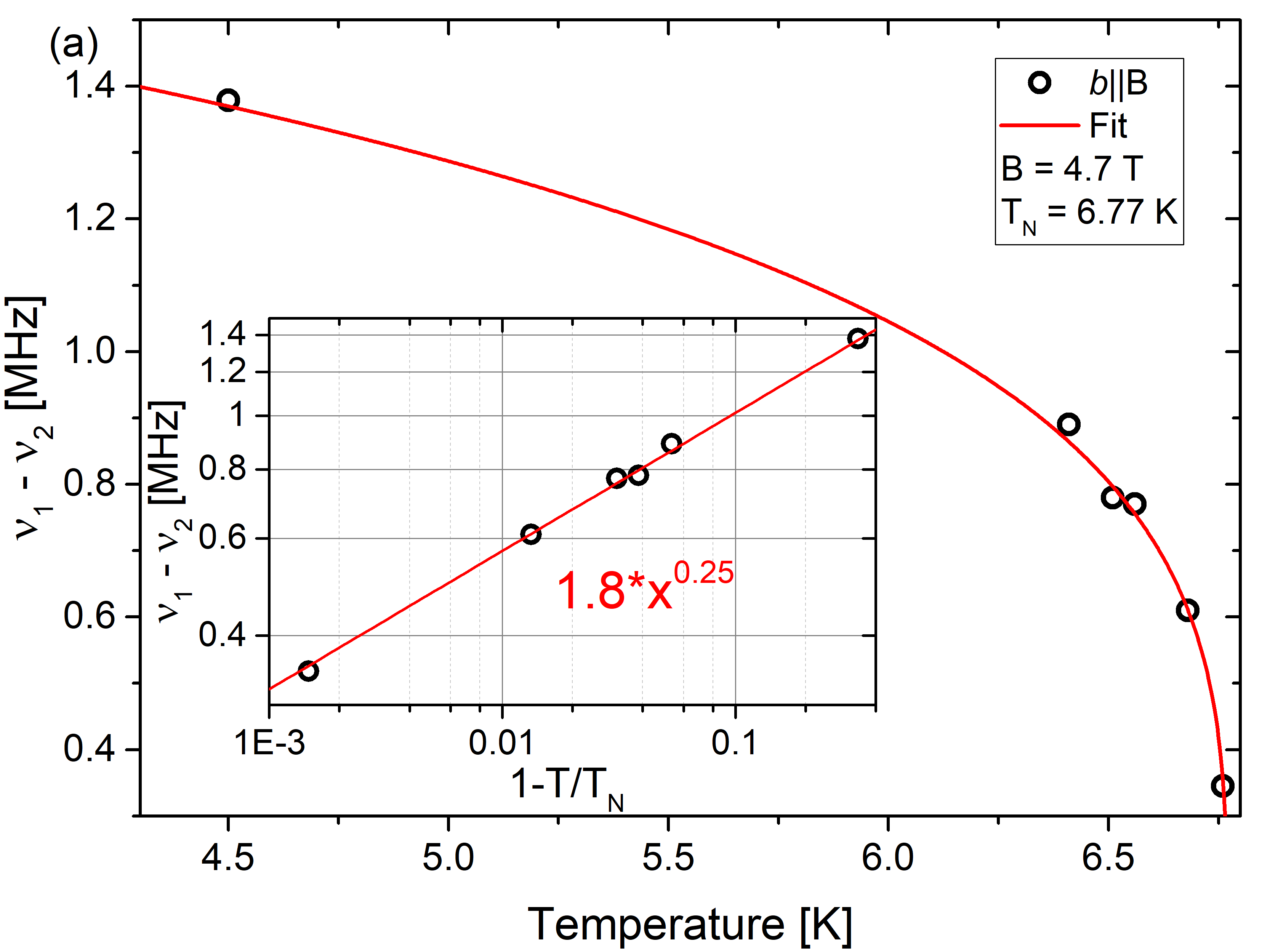} 
	\end{subfigure}%
	
	\begin{subfigure}[b]{0.4\textwidth}
	\includegraphics[width=1\textwidth]{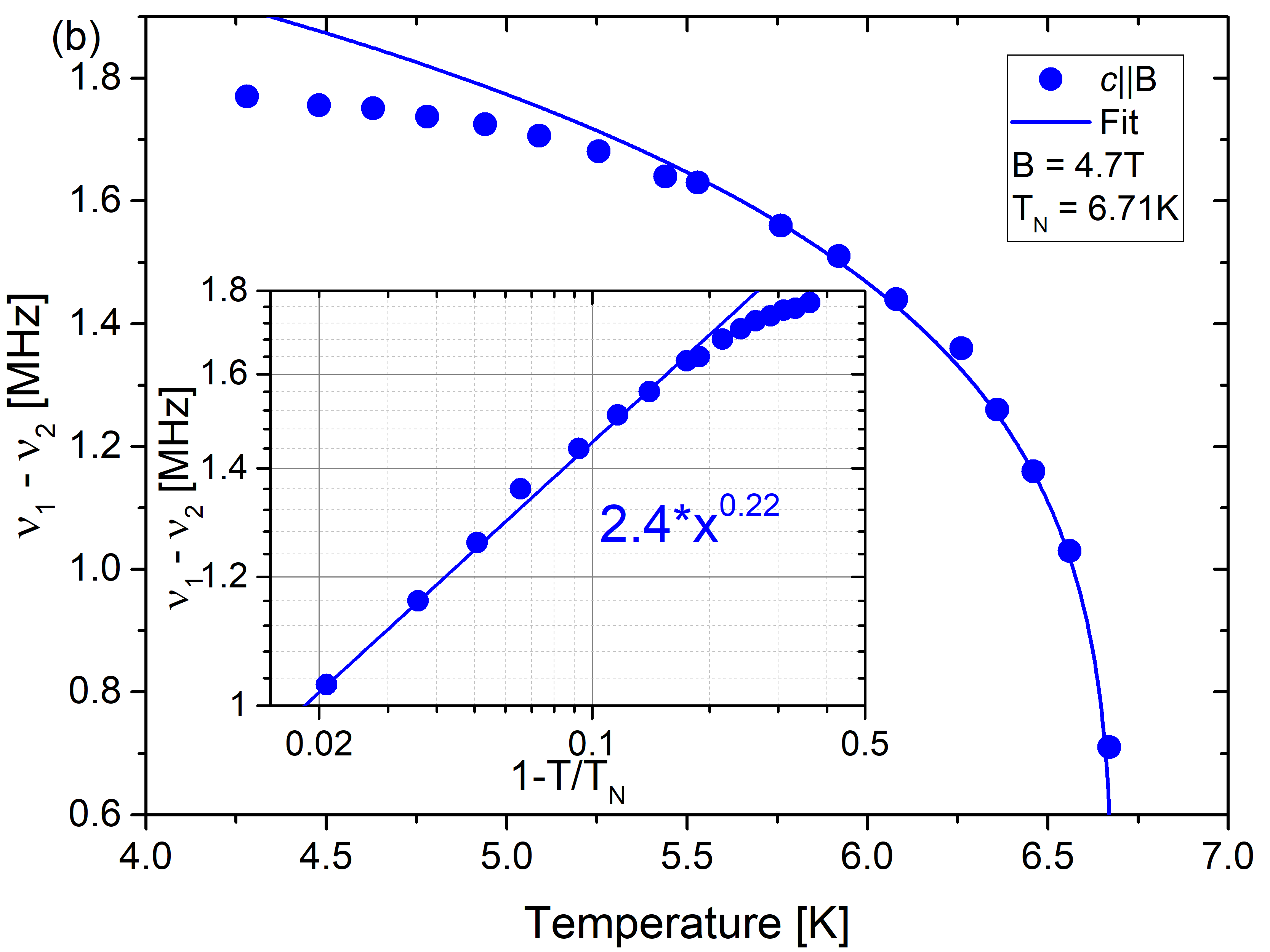} 
	\end{subfigure}
\caption{Temperature dependence of the $^{31}$P line splitting $\Delta F$ for (a) $B\parallel$[010] and (b) $B\parallel$[001] below $T_N$. Insets: fits of $\Delta F(T)$ using Eq.~(\ref{power}) plotted versus x~=~$1-\frac{T}{T_N}$.}
\label{fig:Tn}
\end{figure}

\subsection{Orientation of the $^{31}$P NMR Knight shift tensor in PbTCPO} \label{subsec:KnightOrientation}

\begin{figure}
	\begin{center}
		\includegraphics[width=0.5\textwidth]{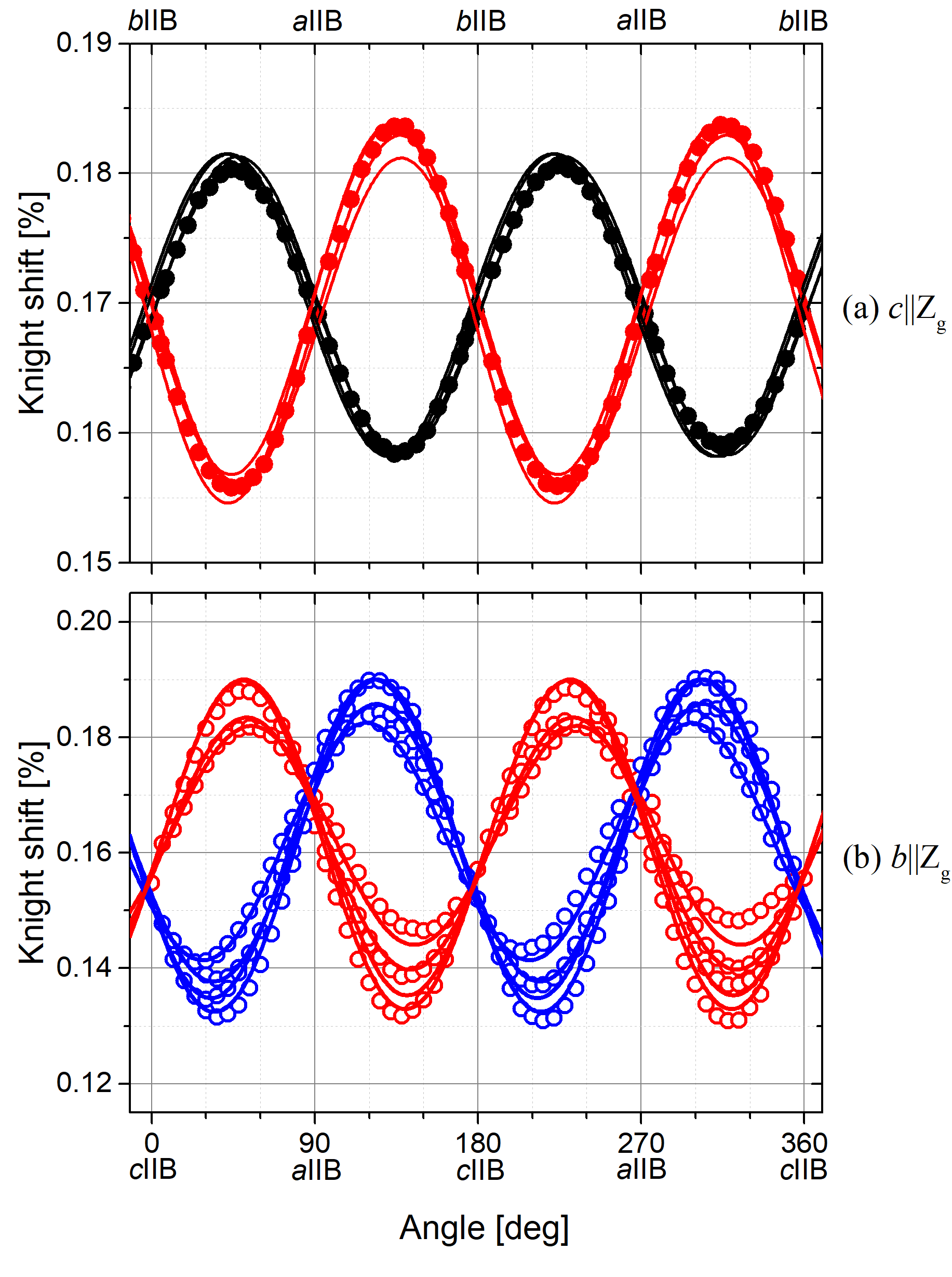} 
		\caption{$^{31}$P NMR measured at room temperature by rotating PbTCPO single crystal around (a) $c$ axis [001] and (b) $b$ axis [010]. The experimental results are fitted using Euler transformations from  PAS to the laboratory frame.}
		\label{fig:PTCP_rot}
	\end{center}
\end{figure}

By rotating the single crystal, it is possible to acquire the exact alignment of the anisotropic Knight shift tensors at the locations of the observable nuclei. PbTCPO single crystal rotation was performed with the same technique as in our work for BaTCPO \cite{Rasta2020} at room temperature and an external magnetic field $B_\text{ext} = 4.7~\text{T}$. The results of sample \#2 are presented in Fig.~\ref{fig:PTCP_rot}. The rotation was performed around two crystallographic axes: $c$ axis ([001]) and $b$ axis ([010]), where [100] and [010] are symmetry equivalent directions. The axis of rotation was placed perpendicular to the external magnetic field $B_\text{ext}$ and the $^{31}$P NMR spectrum was gathered after every $6^{\circ}$ for a full $360^{\circ}$ turn.

The principal values of the Knight shift tensor were found by measuring the static PbTCPO powder sample (not shown): $ K_{11}=1.93 $\%, $ K_{22}=1.75 $\%, $ K_{33}=1.26 $\%. The diagonal shift tensor is transformed into different frames of reference: principal axis system (PAS) $\rightarrow$ crystal frame (CF) $\rightarrow$ goniometer frame (GF) $\rightarrow$ laboratory frame (LF). A ZYZ Euler angle rotation matrix conducts each transformation \cite{easyspin_euler} $R(\alpha,\beta,\gamma)$, where the new system of reference $K'$ is acquired by $K'=RKR^{\mathrm T}$. The LF is aligned with the direction of $B_\text{ext}$, so the transformation from GF to LF is defined by the angle of rotation $R(0,90,x)$, where $ x $ is the angle of rotation. The transformation from CF to GF is defined by the initial position of the crystal onto the goniometer: $ R(90,90,0) $ for  Z$_g\parallel$\textit{b} and $ R(0,0,90) $ for Z$_g\parallel$\textit{c}. PbTCPO crystallizes in the chiral tetragonal space group $P42_12$. Crystallographically there is one symmetry-unique P site, and the space-group operations generate eight P positions in the unit cell. Accordingly, the local $^{31}$P Knight-shift tensors at these sites are not independent, but are symmetry-related images of the same local tensor in the corresponding local environments. Thus, the eight PAS$\rightarrow$CF transformations are constrained by crystal symmetry, after which the common CF$\rightarrow$GF$\rightarrow$LF transformations are applied for the given goniometer mounting.

An exhaustive grid search was performed to determine the PAS$\to$CF ZYZ Euler angles $R(\alpha,\beta,\gamma)$ that reproduce the $^{31}$P rotation data. We sampled $\alpha\!\in\![-90^\circ,90^\circ]$, $\beta\!\in\![0^\circ,180^\circ]$, and $\gamma\!\in\![0^\circ,360^\circ]$ with $1^\circ$ steps. For each $(\alpha,\beta,\gamma)$, the diagonal PAS tensor was propagated to the LF for the two GF mountings (about $[010]$ and about $[001]$), yielding model curves
\[
K^{\rm calc}_{b}(\phi;\alpha,\beta,\gamma),\qquad
K^{\rm calc}_{c}(\phi;\alpha,\beta,\gamma).
\]
The experimental rotation curves were first fit by cosine templates $K^{\rm fit}_{b,\ell}(\phi)$ (eight $[010]$ lines, $\ell=1,\dots,8$) and $K^{\rm fit}_{c,m}(\phi)$ (two $[001]$ lines, $m\in\{1,2\}$). Pointwise residuals were defined as
\begin{align*}
r^{(b)}_{i,\ell}&=K^{\rm fit}_{b,\ell}(\phi_i)-K^{\rm calc}_{b}(\phi_i;\alpha,\beta,\gamma),\qquad \\
r^{(c)}_{j,m}&=K^{\rm fit}_{c,m}(\phi_j)-K^{\rm calc}_{c}(\phi_j;\alpha,\beta,\gamma),
\end{align*}
with $i=1,\dots,N_b$ and $j=1,\dots,N_c$ indexing the sampled angles of the $[010]$ and $[001]$ scans, respectively. These were reduced to angle–averaged RMS residuals
\[
R_{b,\ell}=\Bigl[\tfrac{1}{N_b}\sum_i \bigl(r^{(b)}_{i,\ell}\bigr)^2\Bigr]^{1/2},\qquad
R_{c,m}=\Bigl[\tfrac{1}{N_c}\sum_j \bigl(r^{(c)}_{j,m}\bigr)^2\Bigr]^{1/2},
\]
with equal weighting (all residuals constructed identically).

Because only one of the two $[001]$ lines belongs to the same crystallographic $^{31}$P site as a given $[010]$ line, each $[010]$ line $\ell$ was paired with the better–matching $[001]$ line and a per–site scalar was formed:
\[
\mathcal{R}_\ell(\alpha,\beta,\gamma)=
\Biggl[\frac{R_{b,\ell}^2+\min_{m\in\{1,2\}}R_{c,m}^2}{2}\Biggr]^{1/2}.
\]
Filling the ZYZ grid with $\mathcal{R}_\ell(\alpha,\beta,\gamma)$ at each sampled $(\alpha,\beta,\gamma)$ produces eight $361\times181\times181$ 3D arrays (one grid per P site), where each voxel stores the corresponding RMS residual. Candidate solutions were obtained by visually confirming the number of low-residual clusters, applying \textit{k-means} ($k=4$ in our case), and selecting the minimum residual in each cluster. For every site we observe a \emph{pair} of minima with indistinguishable $\mathcal{R}_\ell$ yet distinct Euler triples; these are symmetry–equivalent under tetragonal $C_4$ operations (together with the sign/permutation freedoms of a second–rank tensor) and thus produce identical $K(\phi)$ curves.

The principal values of the Knight shift tensor vary slightly from the powder spectrum due to small temperature variations and orientation-dependent demagnetization (Lorentz) fields in the crystal. The exact positioning of the single crystal on the goniometer also introduces a few degrees of uncertainty in the rotation into the GF. In near regions of the Knight shift tensor ($K$) principal values and the ZYZ Euler angle set ($R_b$) that turns the axis system into the GF, the minimum points of the eight local minima were refined using \textit{steepest-ascent hill-climbing} method. Variations and combinations of $R_b$ and $K$ were explored by comparing the RMS of the eight PAS-to-CF angle-set residuals. The best result was obtained with the following parameters: $R_b = R(0, 90, 3); K = (K_{11}, K_{22}, K_{33}) = (1.9006, 1.80059, 1.2248)$. 

\begin{figure}[ht!]
	\includegraphics[width=0.4\textwidth]{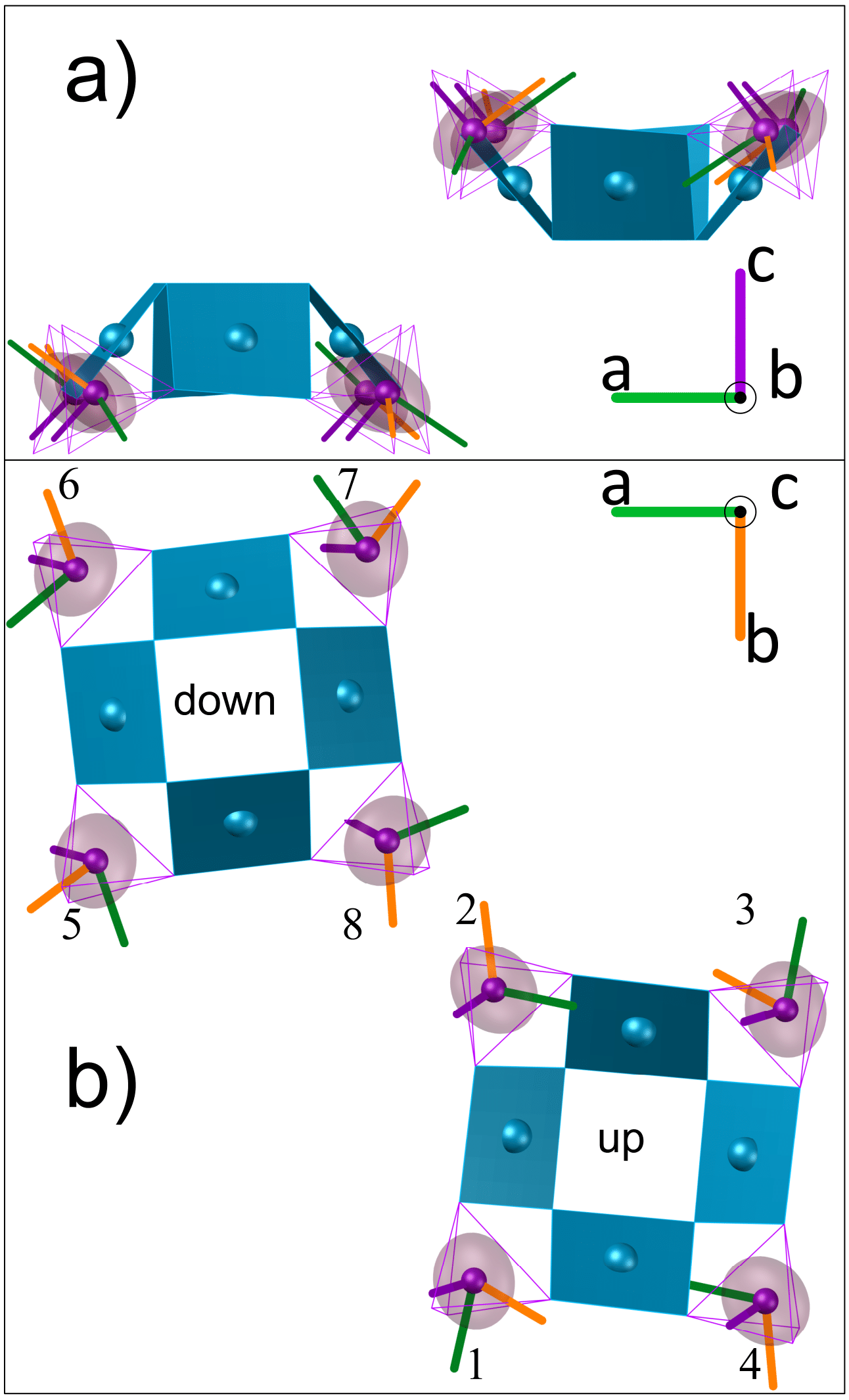} 
	\caption{The configurations of magnetic tensors at temperature \textit{T} = 295 K resulting from $^{31}$P NMR of rotating the single crystal in the (a) [001] view and (b) [010] view. There are two distinct square cupolas where the tensors are represented by three principal vectors - green, orange, dark-violet, which correspond to the $K_{PAS}$ values $K_{11}$, $K_{22}$, $K_{33}$. The defined vector sizes are rotated.}
	\label{fig:PbTCPO_295K}
\end{figure}

The resulting angle set is presented in Table~\ref{295K_rot_results}. These transformations can be used to visualize the Knight shift tensors in the crystal. In Fig.~\ref{fig:PbTCPO_295K}, we propose one possible configuration by pairing four tensors through the four $\gamma$ angles exhibiting 90$^\circ$ $C_4$ symmetry and positioning them on the up and down cupolas according to the $\beta$ angle values. We note that the spatial arrangements of the eight are not fixed by experiment and may be rearranged according to physical considerations.

\begin{table}[h]
	\caption{Euler ZYZ angles for transforming PbTCPO single crystal from the principal axis system into the eight equivalent P positions in the crystal frame. }
	\label{295K_rot_results}
	\begin{ruledtabular}
	\begin{tabular}{c  c c c}
		No. & $\alpha$ & $\beta$ & $\gamma$\\ \hline
		1  & 17  & 43 & 52   \\ 
		2  & 34  & 45 & 144  \\ 
		3  & 17  & 43 & 234   \\ 
		4  & 34  & 43 & 323   \\ 
		5  & -15 & 137 & 47   \\ 
		6  & -14 & 135 & 136   \\ 
		7  & -2  & 137 & 224   \\ 
		8  & -28 & 136 & 319   \\
	\end{tabular}
\end{ruledtabular} 
\end{table}

\subsection{$^{31}$P spin--lattice relaxation time $T_1$}
\begin{figure}[h!]
	\includegraphics[width=0.4\textwidth]{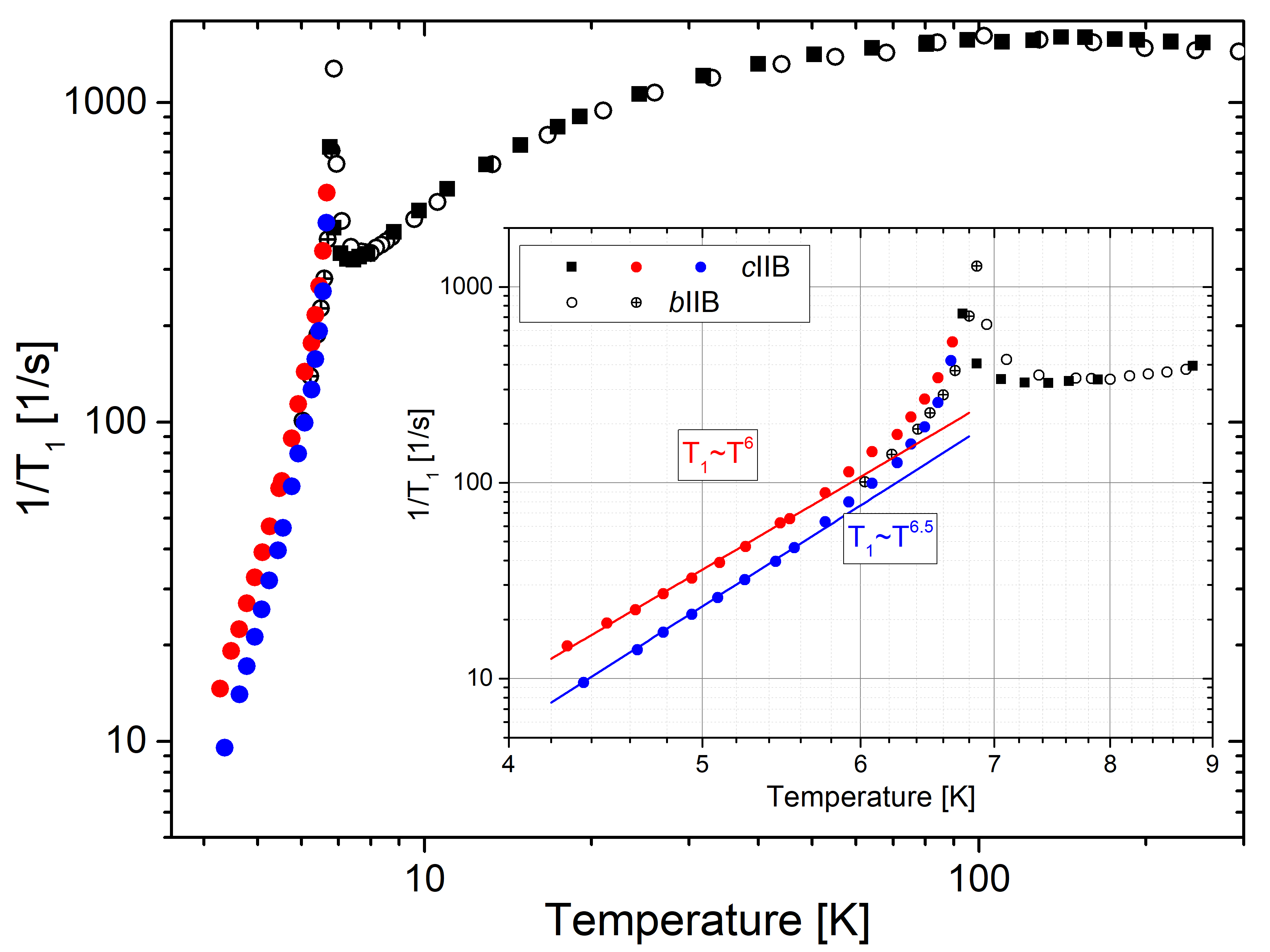}
	\caption{\label{PTCP_295K} Temperature dependence of the $^{31}$P spin--lattice relaxation rate $1/T_1$ of PbTCPO for $\textit{B}$$\parallel[001]$ and $\textit{B}\parallel[010]$. The inset shows a log--log plot of $1/T_1$ versus $T$ highlighting a power-law decrease below $T_N$.}
\end{figure}

The $^{31}$P spin--lattice relaxation rate was measured by the inversion--recovery method with $\textit{B}\parallel[010]$ and $\textit{B}\parallel[001]$ (Fig.~\ref{PTCP_295K}). For $^{31}$P ($I=1/2$), the recovery is well described by a single exponential over the full temperature range,
\begin{equation}
	M(\tau)=M_0\!\left[1-A\exp(-\tau/T_1)\right],
\end{equation}
where $M(\tau)$ is the nuclear magnetization at delay time $\tau$, $M_0$ is the equilibrium magnetization, $A\lesssim 2$ reflects the inversion-pulse efficiency, and $T_1$ is the spin--lattice relaxation time.

For $T\gtrsim 60$~K, $1/T_1$ becomes weakly temperature dependent, consistent with the exchange-narrowed paramagnetic regime where relaxation is driven by fast fluctuations of the local hyperfine field. In this limit, Moriya's theory gives an estimate for the high-$T$ relaxation plateau \cite{Moriya1956},
\begin{equation}
	\frac{1}{T_1}=\frac{2\gamma_N^2\sqrt{2\pi}\,S(S+1)}{3\,\omega_E\,z'}\,H_{\rm hf}^2.
\end{equation}
Here $\gamma_N$ is the nuclear gyromagnetic ratio, $S=1/2$ is the Cu$^{2+}$ electronic spin, and $H_{\rm hf}$ is the (nearly isotropic) transferred hyperfine coupling extracted from the $K$--$\rchi$ analysis (in kOe$/\mu_B$). The prefactor 2 accounts for the two transverse components of the fluctuating hyperfine field that drive relaxation. We take $z'=4$, corresponding to four nearest Cu spins coupled to a given P site through the Cu--O--P network.
The exchange frequency is written as
\begin{equation}
	\omega_E=\frac{|J|k_B}{\hbar}\sqrt{\frac{2zS(S+1)}{3}},
\end{equation}
with $z=2$ the number of dominant in-cupola Cu neighbors associated with the leading exchange scale $J$.

Using the high-$T$ plateau values $1/T_1\simeq 1440$~s$^{-1}$ for $\textit{B}\parallel[010]$ and $1540$~s$^{-1}$ for $\textit{B}\parallel[001]$, we obtain $J/k_B\simeq 8.9$~K and $7.0$~K, respectively. We therefore quote an average exchange scale $J/k_B=8.0(9)$~K, where the uncertainty reflects the orientation spread (Table~\ref{tab:internal}). This estimate is of the same order as expected from the Curie--Weiss scale and serves as an effective comparison for the dominant in-cupola exchange interactions. We admit that in the square cupola compounds the spin structure involves a number of spin-spin interactions \cite{Islam2018,Testa2022} and the NMR relaxation estimates only effective fluctuation rates of the local magnetic field at frequencies close to the Larmor frequency.

On cooling toward $T_N$, the slowing down of Cu$^{2+}$ spin fluctuations enhances the low-frequency spectral weight at the nuclear Larmor frequency, resulting in a pronounced peak in $1/T_1$ at the transition. The amplitude and width of this critical enhancement depend on how the transferred Cu--O--P hyperfine coupling weights the Cu-spin fluctuations, and on the effective dimensionality of the critical correlations near $T_N$.

Below $T_N$, $1/T_1$ decreases rapidly as static order develops and the remaining relaxation is governed by low-energy spin excitations that modulate the local hyperfine field. The observed power-law behavior, $1/T_1\propto T^{6}$--$T^{6.5}$ for $B \parallel c$ (depending on the resolved $^{31}$P site), is consistent with a two-magnon Raman relaxation process expected for insulating antiferromagnets \cite{Abragam1961}. Compared with the $T^{7}$ behavior reported for BaTCPO \cite{Rasta2020} and SrTCPO \cite{Islam2018}, the slightly smaller exponent and the stronger critical peak in PbTCPO point to modified low-energy spin dynamics within the $A$TCPO family, which can be associated with differences in interlayer coupling set by the $A$-site chemistry and quadrupole stacking.

\subsection{$^{31}$P NMR with rotating crystal at $T$ = 4.5 K}

\begin{figure}[htbp]
	\includegraphics[width=0.45\textwidth]{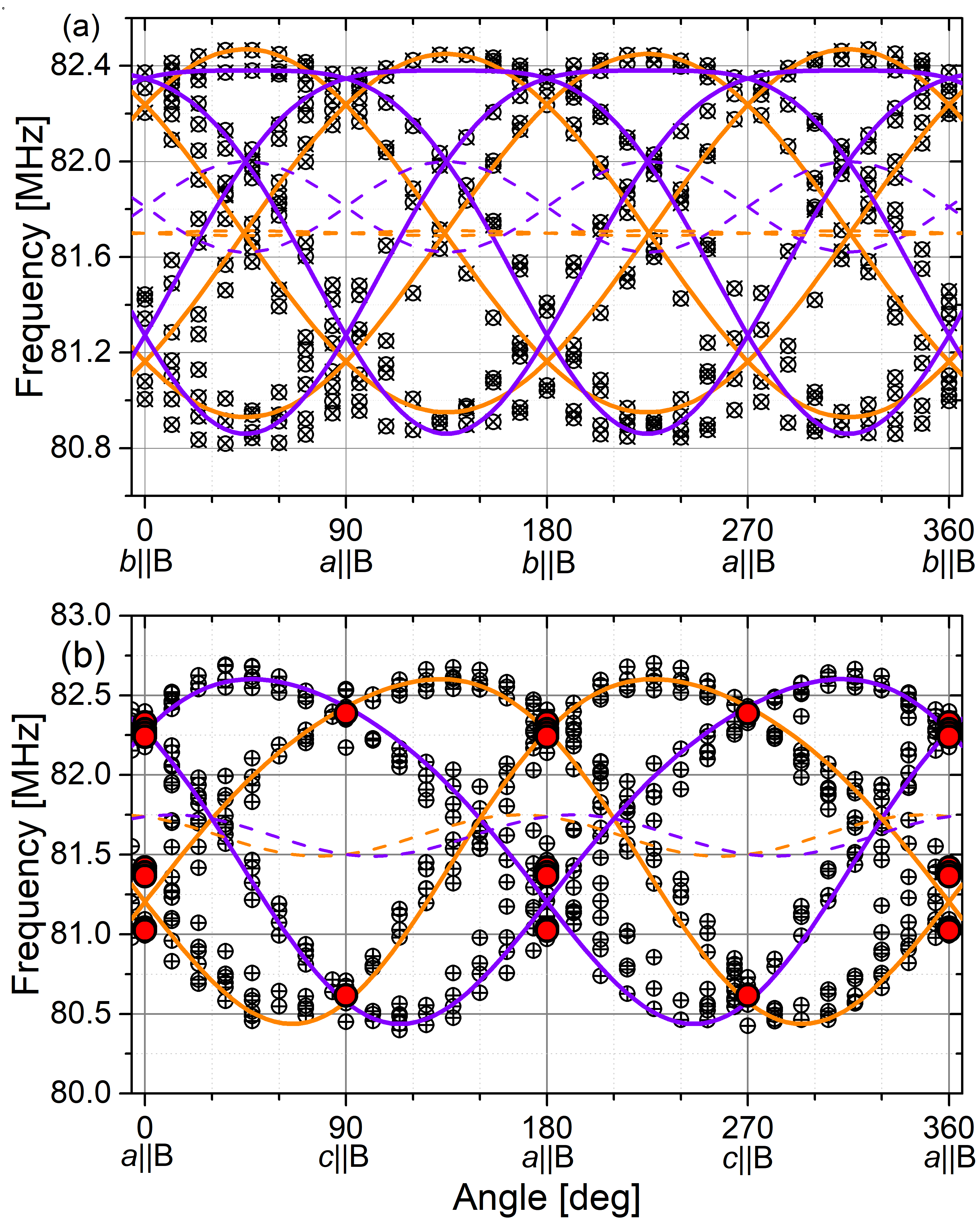} 
	\caption{$^{31}$P resonance frequencies of PbTCPO single crystal at temperature $T$ = 4.5 K and $B_\text{ext} = 4.7$~T by rotating 360$^\circ$ around (a) $Z_g$$\parallel$[001] and (b) $Z_g$$\parallel$[010]. Experimental results are approximated with Eq.~(\ref{lahendus}) using data in Table~\ref{tab:4K}. The red points on (b) visualize the result from Fig.~\ref{fig:Knight}.}
	\label{rot_4K}
\end{figure}

The resonance-frequency angle dependence of rotating the single crystal around $Z_g$$\parallel$[001] and $Z_g$$\parallel$[010] was measured at $T = 4.5 $~K. Both measurements were conducted for two distinct samples. A complex angle dependence at room temperature suggests the presence of different domains and/or chirality in the first (\#1) sample. The second (\#2) sample was confirmed to be monodomain with levorotary chirality. At low temperature, both samples exhibit an identical $^{31}$P NMR angle dependence (Fig.~\ref{rot_4K}). Rotating the sample reveals lines for each of the eight phosphorus ions in the unit cell. The angle dependence is well described by the sum of the projection of the static internal field onto the external field and the contribution from the second–rank Knight–shift tensor,
\begin{equation} \label{lahendus}
F(\alpha)=K+L\cos(\alpha-\alpha_1)+M\cos(2(\alpha-\alpha_2)),
\end{equation}
where \(K\) is an offset equal to the carrier (Larmor) frequency plus the isotropic shift, the second term describes the angle dependence from the projection of the local static field onto the external field direction, and the third term is the angle dependence produced by the second–rank Knight–shift tensor. The fitting data is presented in Table \ref{tab:4K}.

The rotation about $Z_g\parallel[001]$ [Fig.~\ref{rot_4K}a] yields two quartets of lines, consistent with the eight crystallographic P sites. Within each quartet, the phase offsets are close to $90^\circ$, while the two quartets differ by a small offset in the second-harmonic term ($M$), reflecting the Knight-shift anisotropy. The rotation about $Z_g\parallel[010]$ [Fig.~\ref{rot_4K}b] yields four resolved branches, grouped into two pairs with a relative phase shift, again consistent with the site symmetry captured by Eq.~(\ref{lahendus}). Small deviations from the fit may reflect the proximity to $T_N$, where the ordered moment and/or domain populations are still evolving with temperature. It is seen that the second-rank Knight shift (tensor) contribution, typically dominant in the paramagnetic state, is rather large. The Knight shift data in Fig.~\ref{fig:Knight} show peaks at 81.40 and 81.04~MHz for $b\parallel B$, which are not fully resolved at this scale; however, for the goal of evaluating the strength of the internal field, this is sufficient. 

\begin{table}[tbp]
	\vskip 15pt
	\caption{Fitting parameters $K$, $L$, $M$, $\alpha_1$, and $\alpha_2$ of the $^{31}$P NMR rotation patterns, according to Eq.~(\ref{lahendus}).}
	\label{tab:4K}
	\begin{ruledtabular}
	\begin{tabular}{c  c c c c c}
		No. & $K$ & $L$ &$\alpha_1$ & $M$ & $\alpha_2$ \\ [1ex] 
		\hline
		\multicolumn{6}{ c }{Rotation around the $c$ axis}\\
		1 & 81.70 & 0.76 & 45  & 0.01 & 45\\ 
		2 & 81.70 & 0.76 & 135 & 0.01 & 45\\ 
		3 & 81.70 & 0.76 & 225 & 0.01 & 135\\
		4 & 81.70 & 0.76 & 315 & 0.01 & 135\\ 
		5 & 81.81 & 0.76 & 45  & 0.19 & 135\\  
		6 & 81.81 & 0.76 & 135 & 0.19 & 45\\ 
		7 & 81.81 & 0.76 & 225 & 0.19 & 135\\
		8 & 81.81 & 0.76 & 315 & 0.19 & 45\\  [1ex]
		\multicolumn{6}{ c }{Rotation around the $b$ axis}\\
		1 & 81.62 & 1.07 & 60 & 0.13 & 168\\ 
		2 & 81.62 & 1.07 & 120 & 0.13 & 12\\ 
		3 & 81.62 & 1.07 & 240 & 0.13 & 168\\ 
		4 & 81.62 & 1.07 & 300 & 0.13 & 12\\ [1ex]
	\end{tabular}
\end{ruledtabular} 
\end{table}
\begin{figure}
	\includegraphics[width=0.25\textwidth]{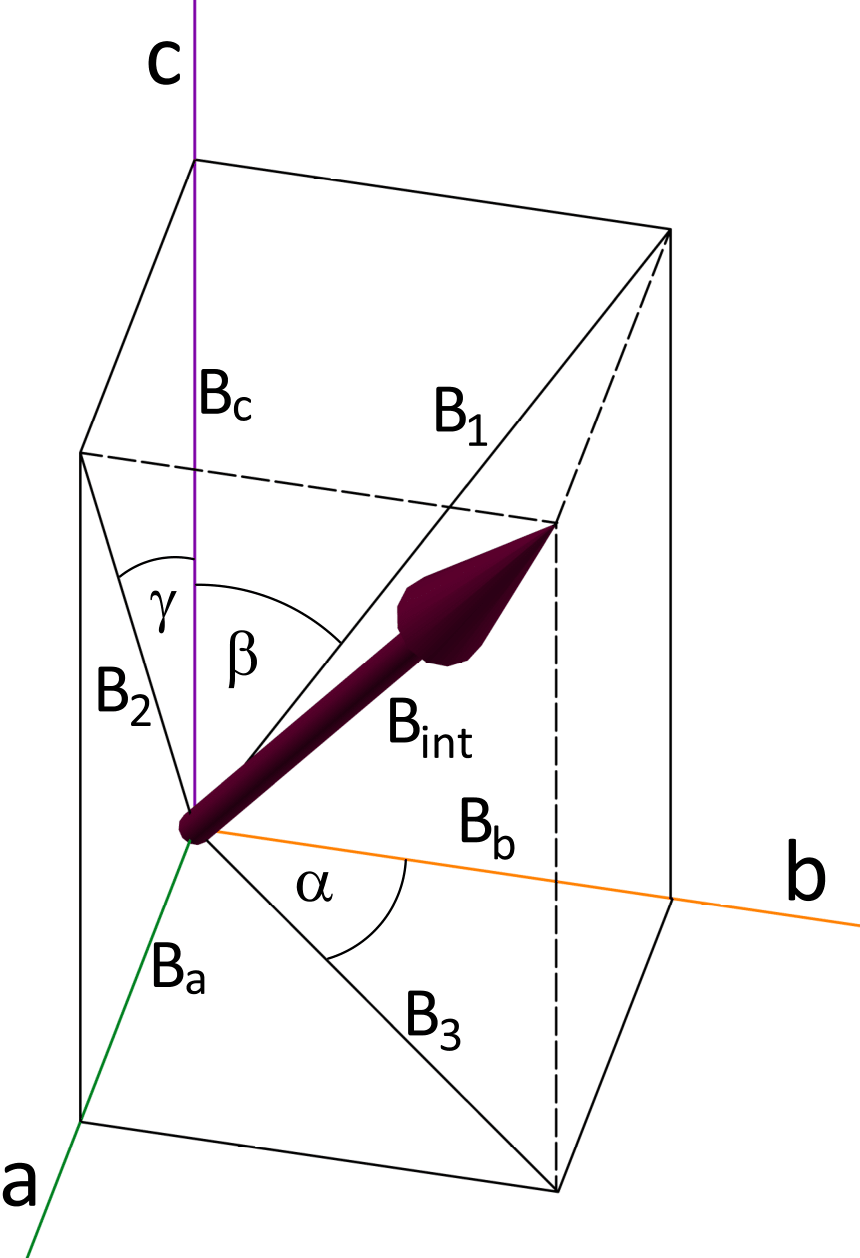} 
	\caption{ Scheme of the local field direction on a phosphorus ion in the PbTCPO single crystal at temperature $T$ = 4.5 K. \textit{B}$_1$,~\textit{B}$_2$, and \textit{B}$_3$ represent the projections of B$_\text{int}$ into the \textit{bc}, \textit{ac}, and \textit{ab} planes, respectively that are directly seen from the amplitudes of $^{31}$P NMR experiments. \textit{B}$_a$,~\textit{B}$_b$, and \textit{B}$_c$ are the direction cosines that are used in constructing a 3D view in order to find the resulting B$_\text{int}$.}
	\label{axes}
\end{figure}

The fitted amplitudes $L$ directly yield the magnitudes of the local hyperfine field projections, from which the vector components of $B_\text{int}$ can be reconstructed as shown in Fig.~\ref{axes}. The cosine amplitudes $L$ correspond to the local field projections $B_1,B_2$, and $B_3$ and using the gyromagnetic ratio of $^{31}$P $\gamma/2\pi = 17.237$ MHz/T, we get $B_3 = 44.1$ mT, $\alpha = \pm45^\circ$ from $Z_g$$\parallel$\textit{c}, and $B_1=B_2=62.1$ mT, $\beta = \gamma =\pm30^\circ$ from $Z_g$$\parallel$\textit{b}. The projections onto the axes are then $B_a = B_b = 31.1$ mT, $B_c = 53.8$ mT, and for the internal field we get $B_\text{int} = 69.5$ mT. In the work of Ihara \textit{et al.}~\cite{Ihara2025a} $B_\text{int} = 71.3$ mT was found at a lower temperature of \textit{T} = 1.5 K, which is approximately twice that of the antiferroically stacked analogues. 

In the ordered state the internal field at $^{31}$P can be decomposed as $\textit{B}_{\rm int}=\textit{B}_{\rm dip}+\textit{B}_{\rm tr}$, where $\textit{B}_{\rm dip}$ is the classical dipolar field from the ordered Cu moments and $\textit{B}_{\rm tr}$ is the transferred hyperfine field generated through the Cu--O--P network. Our calculated dipolar-fields are $\textit{B}_{\rm dip} = 37.9$~mT for PbTCPO and $\textit{B}_{\rm dip} = 51.6$~mT for BaTCPO. Those calculated values are 16~mT larger than $\textit{B}_{\rm int}$ for BaTCPO and 32.1~mT smaller than $\textit{B}_{\rm int}$ for PbTCPO, respectively.

A key point is that $\textit{B}_{\rm dip}$ at an off-site ligand nucleus does not simply follow the net magnetization, but is set by a geometry-dependent dipolar tensor. Observing layer-resolved dipolar sums, we find that contributions from the nearest $c$-translated layers point opposite to the dominant on-layer contribution for the ferroic reference configuration, producing partial cancellation in the vector sum. When the stacking along $c$ is antiferroic, the alternating sign of the ordered moments converts the cancellation into an additive effect, and therefore yields a larger $\textit{B}_{\rm dip}$ than in the ferroic case despite the antiferroic stacking of the magnetic-quadrupole pattern.

The fact that $\textit{B}_{\rm dip}$ exceeds $\textit{B}_{\rm int}$ in BaTCPO implies that the transferred contribution opposes and partially compensates the dipolar field in the ordered state. The much larger $\textit{B}_{\rm int}$ in PbTCPO compared with BaTCPO indicates that the net local field is enhanced not only by the stacking-controlled interlayer cancellation, but also by the transferred component. A complete microscopic description would benefit from first-principles hyperfine-tensor calculations (DFT+U/hybrid) and from fitting an effective transferred-hyperfine model to our site-resolved ordered-state $^{31}$P internal-field data. 

\subsection{$^{63,65}$Cu ZFNMR in liquid-He at $T$ = 4.2~K} \label{subsec:CuZFNMR}

\begin{figure}
	\centering
	\includegraphics[width=0.4\textwidth]{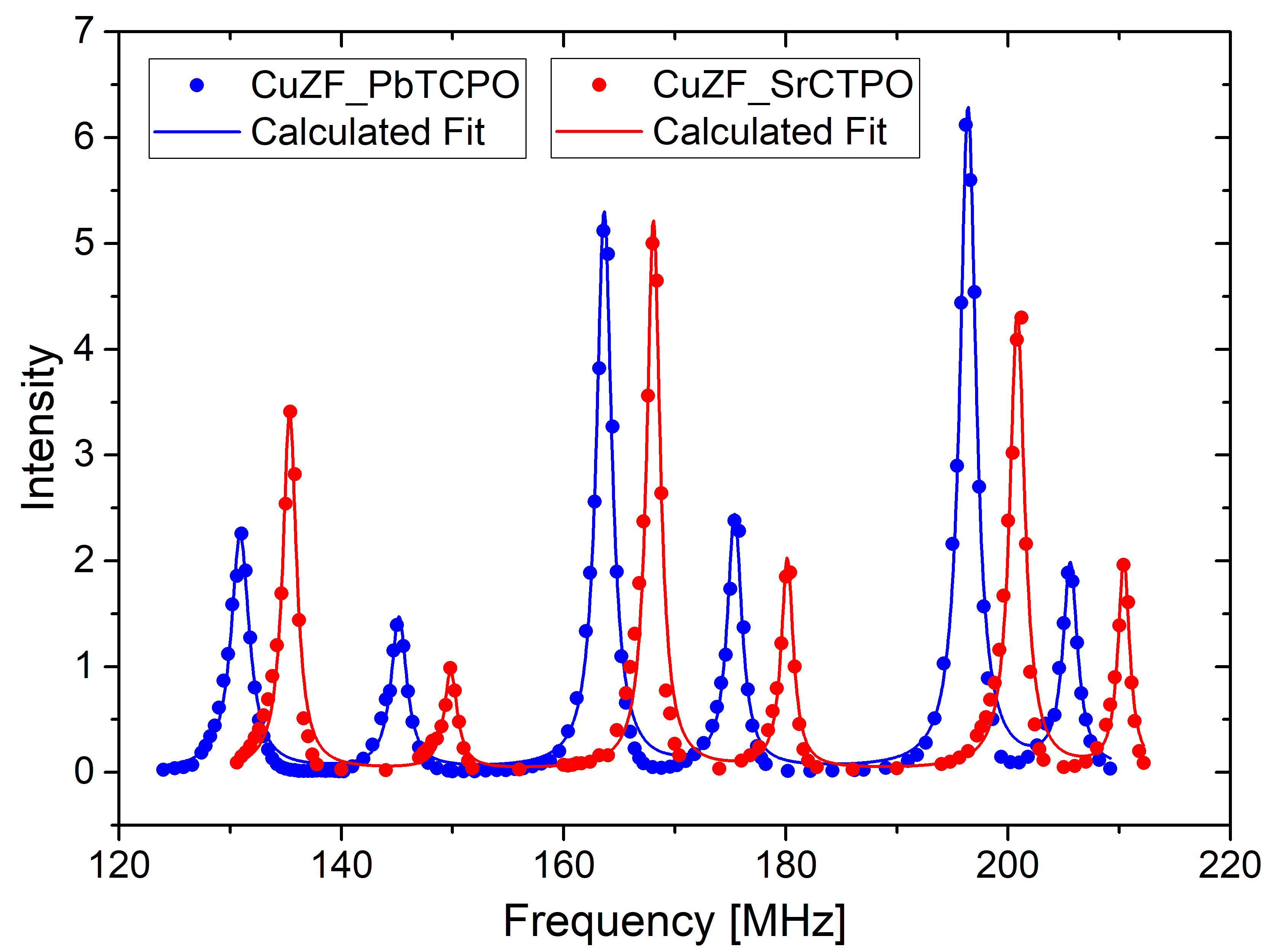} 
	\caption{ $^{63,65}$Cu ZFNMR spectrum of PbTCPO (blue) and SrTCPO (red) at $T$ = 4.2 K. The dots represent experimental data, and the line corresponds to a Lorentzian fit, obtained by the transition frequencies obtained from Eq.~(\ref{eq:HQ_tensor_fieldframe}), using $\nu_L$ and $\nu_Q$ values from Table~\ref{tab:ZFNMR}.}
	\label{fig:ZFNMR}
\end{figure}

$^{63,65}$Cu Zero-field NMR (ZFNMR) directly probes the internal field at the Cu sites and the local EFG tensor in the magnetically ordered state. The two copper isotopes $^{63}$Cu and $^{65}$Cu both have nuclear spin $I = 3/2$, and hence exhibit a central transition ($m_I = \pm \tfrac{1}{2} \leftrightarrow \mp \tfrac{1}{2}$) and two satellite transitions ($m_I = \pm \tfrac{3}{2} \leftrightarrow \pm \tfrac{1}{2}$). In zero applied field the splitting of these transitions is governed by the competition between the strong internal Zeeman field from the ordered Cu$^{2+}$ moments and the quadrupolar interaction with the local EFG. Representative spectra for PbTCPO and SrTCPO are shown in Fig.~\ref{fig:ZFNMR} (BaTCPO was presented in \cite{Rasta2020}); the six lines (three per isotope) are well resolved and can be followed across the $A$TCPO series.

\begin{table}[ht!]
	\vskip 15pt
	\renewcommand{\arraystretch}{1.5}
	\centering
	\caption{Parameters of $A$TCPO at temperatures below $T$ = 5 K from $^{63,65}$Cu NMR at $B_\text{ext}$ = 0 T and $^{31}$P NMR at $B_\text{ext}$ = 4.7 T.}
	\label{tab:internal}
	\begin{ruledtabular}
	\begin{tabular}{c c c c c}
		nucleus & sample & $H_\text{int}$ (T) &$\theta$ (deg) & $\eta$ \\ 
		\hline
		$^{63,65}$Cu & BaTCPO & 14.76(3) & 8 & 0.73\\  [1ex] 
		& PbTCPO & 14.50(6) & 0 & 0.43 \\  
		& SrTCPO & 14.90(6) & 2 & 0 \\ \hline 
		& & $H_\text{int}$ (mT) &$J/k_B$ (K) \\ \hline
		$^{31}$P & BaTCPO & 35.6(20) & 11.3(5) \\ 
		& PbTCPO & 69.5(9) & 8.0(9) \\ 
		& SrTCPO & 34.6(2)$^\mathrm{a}$ & 8.0(2) \\ 
	\end{tabular}
\end{ruledtabular}
\renewcommand{\arraystretch}{1}
\footnotetext[1]{From Ref.~\cite{Islam2018}.}
\end{table}

\vspace{4pt}
\noindent\textbf{General Hamiltonian.}
For a spin $I = 3/2$ nucleus in a static internal field $\mathbf{H}_\text{int}$ and an axially asymmetric EFG, the Hamiltonian for isotope $\alpha = 63,65$ can be written in the EFG principal-axis system (PAS) as~\cite{Mehring1983}
\begin{equation} \label{eq:Htotal_general}
	\hat{\mathcal H}^{(\alpha)} =
	-\gamma_\alpha \hbar\,\mathbf{I}\cdot\mathbf{H}_\text{int} + \frac{h\nu_Q^{(\alpha)}}{6}
	\Big[
	3\hat I_z^2 - I(I+1) + \eta\big(\hat I_x^2 - \hat I_y^2\big)
	\Big].
\end{equation}
The first term represents the Zeeman interaction between the nuclear magnetic moment $\boldsymbol{\mu}_\alpha = \gamma_\alpha\hbar\,\mathbf{I}^\mathsf T$ and the internal magnetic field $\mathbf{H}_\text{int}$. The second term is the nuclear quadrupolar interaction, written in terms of
the quadrupole frequency  $\nu_Q^{(\alpha)}$, the EFG asymmetry parameter $\eta$, and the spin operators $\hat{\mathbf I} = (\hat I_x,\hat I_y,\hat I_z)$. In this frame, $\mathbf V_\text{PAS}=\text{diag}(V_{xx},V_{yy},V_{zz})$, and $\nu_Q^{(\alpha)}$ and $\eta$ are related to the principal components via
\begin{equation} \label{eq:nuQ_eta_def}
	\nu_Q^{(\alpha)} = \frac{3eQ_\alpha V_{zz}}{2I(2I-1)h},
	\qquad
	\eta = \frac{V_{xx} - V_{yy}}{V_{zz}},
\end{equation}
so that $V_{xx} = -\tfrac{1}{2}V_{zz}(1-\eta)$,
$V_{yy} = -\tfrac{1}{2}V_{zz}(1+\eta)$, and
$V_{xx}+V_{yy}+V_{zz}=0$.

\vspace{4pt}
\noindent\textbf{Numerical implementation in the internal-field frame.}
For numerical calculations, we work in the internal-field frame, with the $z$-axis aligned with $\mathbf H_\text{int}$.  In this frame, the Zeeman term is diagonal,
\begin{equation} \label{eq:HZ_fieldframe}
	\hat{\mathcal H}^{(\alpha)}_Z = -\gamma_\alpha \hbar H_\text{int}\,\hat I_z,
\end{equation}
with $H_\text{int} = |\mathbf H_\text{int}|$.  The EFG tensor is then obtained by rotating the PAS tensor into the internal-field frame.  It is convenient to separate the overall scale from the tensor shape and introduce a dimensionless traceless tensor $\tilde{\mathbf V}_\text{PAS}(\eta)$ with eigenvalues proportional to $(-1+\eta,-1-\eta,2)$.  The EFG tensor in the field frame is
\begin{equation}\label{eq:V_rotated}
	\tilde{\mathbf V}(\theta,\phi,\eta) =
	R(\theta,\phi)\,
	\tilde{\mathbf V}_\text{PAS}(\eta)\,
	R^\mathsf T(\theta,\phi),
\end{equation}
where $R(\theta,\phi)$ is the usual rotation matrix that tilts the EFG $V_{zz}$ axis by the polar and azimuthal angles $(\theta,\phi)$ with respect to $\mathbf H_\text{int}$.  In this representation, the quadrupolar Hamiltonian can be written compactly as
\begin{equation} \label{eq:HQ_tensor_fieldframe}
	\hat{\mathcal H}^{(\alpha)}_Q =
	\frac{h\nu_Q^{(\alpha)}}{6}
	\,\hat{\mathbf I}^\mathsf T\,\tilde{\mathbf V}(\theta,\phi,\eta)\,\hat{\mathbf I},
\end{equation}
with $\hat{\mathbf I} = (\hat I_x,\hat I_y,\hat I_z)^\mathsf T$ defined in the internal-field frame.  The total Hamiltonian $\hat{\mathcal H}^{(\alpha)}=\hat{\mathcal H}^{(\alpha)}_Z+\hat{\mathcal H}^{(\alpha)}_Q$ is represented as a $4\times 4$ matrix in the
$\{|m_I\rangle\}$ basis ($m_I = \pm \tfrac{3}{2},\,\pm \tfrac{1}{2}$), and its eigenvalues $E_m^{(\alpha)}$ are obtained by numerical diagonalization.

In the joint analysis of $^{63}$Cu and $^{65}$Cu, we use the fact that both isotopes probe the same internal field and EFG tensor at a given Cu site; the isotope dependence enters only through the nuclear constants $\gamma_\alpha$ and $Q_\alpha$. We therefore parametrize the Hamiltonian by a single Larmor frequency $\nu_L^{(63)}$, a single quadrupole frequency $\nu_Q^{(63)}$, and common geometric parameters $(\theta,\phi,\eta)$, while the $^{65}$Cu frequencies are obtained by the exact scalings
\begin{equation}\label{eq:Larmor_quad_scaling}
	\nu_L^{(65)} = \frac{\gamma_{65}}{\gamma_{63}}\, \nu_L^{(63)},
	\qquad
	\nu_Q^{(65)} = \frac{Q_{65}}{Q_{63}}\,\nu_Q^{(63)},
\end{equation}
when constructing the $^{65}$Cu Hamiltonian.  For each trial parameter set $p = \{\nu_L^{(63)},\nu_Q^{(63)},\theta,\phi,\eta\}$ we build $\hat{\mathcal H}^{(63)}(p)$ and $\hat{\mathcal H}^{(65)}(p)$, diagonalize both matrices, and compute the six transition frequencies $f^{(\alpha)}_{k}(p) = [E^{(\alpha)}_{m}-E^{(\alpha)}_{n}]/h$.

\vspace{4pt}
\noindent\textbf{Residual and error analysis.}
The best-fit parameters $(\nu_L^{(63)},\nu_Q^{(63)},\theta,\phi,\eta)$ are obtained by minimizing the joint $\rchi^2$ of the six $^{63,65}$Cu transition frequencies (three lines per isotope). For each isotope $\alpha=63,65$ and line $k=\mathrm{hi},c,\mathrm{lo}$, the residual is defined in the code as $d_k^{(\alpha)} = f_{k,\mathrm{calc}}^{(\alpha)}-f_{k,\mathrm{obs}}^{(\alpha)}$, where $f_{k,\mathrm{calc}}^{(\alpha)}$ is computed by numerical diagonalization of the full spin-$3/2$ Hamiltonian. The experimental one-standard-deviation uncertainty of each observed peak is estimated from its linewidth and signal-to-noise ratio as $\sigma_k^{(\alpha)}=\mathrm{FWHM}_k^{(\alpha)}/[2\,\mathrm{SNR}_k^{(\alpha)}]$. The weighted misfit is then evaluated using the normalized residuals $z_k^{(\alpha)}=d_k^{(\alpha)}/\sigma_k^{(\alpha)}$ and the joint objective function $\rchi^2=\frac{1}{6}\sum_{\alpha,k}\left(z_k^{(\alpha)}\right)^2$.

Uncertainties of the fit parameters are estimated from the local sensitivity of the calculated transition frequencies to $(\nu_L^{(63)},\nu_Q^{(63)},\theta,\phi,\eta)$ at the $\rchi^2$ minimum. Numerically, this is done by finite-difference variations of the parameters
around the optimum, weighted by the experimental line uncertainties $\sigma_k^{(\alpha)}$ used in $\rchi^2$.

The $^{63,65}$Cu ZFNMR Hamiltonian converged well, and the resulting quadrupolar frequency, Larmor frequency values, and the corresponding transition frequencies are presented in Table \ref{tab:ZFNMR}. The transition frequencies are essentially insensitive to the azimuthal angle $\phi$ because the $\phi$-dependent quadrupolar terms are suppressed by the small factor $\eta \sin^2\theta$ (and higher powers of $\sin\theta$); thus for PbTCPO ($\theta\simeq 0^\circ$) and SrTCPO ($\eta\simeq 0$) the dependence vanishes, while even for BaTCPO ($\theta\approx 8^\circ$) the resulting $\phi$-induced frequency modulation remains well below the linewidth/SNR-limited experimental uncertainty. A 2D heatmap was used to observe the physically plausible range of the parameters $\eta$ and $\theta$. For PbTCPO and SrTCPO, these parameters are within a narrow range, with the best values used for the Hamiltonian calculation presented in Table \ref{tab:internal}. BaTCPO showed results with a good enough fitting only for $\theta$ values between 7 and 9, while $\eta$ had little effect on the residual, ranging from 0.4 to 1 for $\theta = 8$. \\

\subsubsection{Point-charge EFG calculations and Cu $3d$-hole occupancy} \label{subsubsec:EFG}
By separating lattice and on-site contributions, it is possible to find the Cu quadrupole interaction by performing the real-space point-charge calculations of the EFG at the Cu site. We developed the numerical code and followed the established point-charge methodology for oxide cuprates \cite{Shimizu1990,Sternheimer1950,Sternheimer1974}. The crystallographic fractional coordinates were taken from Refs.~\cite{Kimura2016_magneto,Kimura2018Acation,Kimura_inorganic} for BaTCPO, PbTCPO, and SrTCPO, respectively.

In the point-charge model, the lattice contribution to the EFG tensor at the Cu site is obtained as a Coulomb sum,

\begin{equation}
	V^{\rm latt}_{ij} = \sum_n \frac{q_n}{4\pi\varepsilon_0}
	\left[ \frac{3r_{n,i}r_{n,j} - r_n^2\delta_{ij}}{r_n^5} \right],
\end{equation}
where $q_n$ are effective ionic charges and $\mathbf{r}_n$ is the vector from the Cu site to ion $n$. The sum was performed within a sphere of radius $r = 100$~\AA, which is sufficient to ensure convergence.  Diagonalization of $V^{\rm latt}_{ij}$ yields the principal components $(V_{xx},V_{yy},V_{zz})$ and, through the eigenvectors, the orientation of the EFG PAS with respect to the crystallographic frame. For all three compounds, we find that the principal $V_{zz}$ axis is close to the local CuO$_4$ plaquette normal, confirming the magnetic structure model suggested by the neutron diffraction study~\cite{Kimura2016_magneto}.

Using the calculated principal component $V_{zz}^{\rm latt}$, we obtain the lattice quadrupole frequencies for $^{63}$Cu as $\nu_Q^{\rm latt} = 0.77(2)~\text{MHz}$ for BaTCPO and SrTCPO, and  $\nu_Q^{\rm latt} = 0.76(2)~\text{MHz}$ for PbTCPO.  The calculated asymmetry parameters of the lattice EFG tensor are
$\eta_{\rm latt}(\text{Ba}) = 0.05$,
$\eta_{\rm latt}(\text{Pb}) = 0.07$, and
$\eta_{\rm latt}(\text{Sr}) = 0.09$,
with an estimated uncertainty of $\pm 0.01$. The anisotropic on-site Cu $3d_{x^2-y^2}$ hole and covalency with the surrounding oxygen ligands add a sizable \emph{valence} EFG contribution whose symmetry differs from that of the purely ionic lattice. To connect the lattice EFG to the experimentally observed quadrupole splittings, we use the Sternheimer expression
\begin{equation} \label{eq:nuQ}
		\nu_Q \simeq
		\left(1-\gamma_\infty\right)\,\nu_Q^{\rm latt}
		+
		(1-R)\,n_d\,B_\text{free-ion},	
\end{equation}
\\
\noindent
where $n_d$ denotes the occupancy of the Cu $3d_{x^2-y^2}$ hole orbital, i.e. the fraction of one hole residing on the Cu site in the local CuO$_4$ plaquette basis, $(1-\gamma_\infty)$ and $(1-R)$ are the antishielding and valence reduction factors, and $B_{\rm free\text{-}ion}$ is the EFG of a fully localized Cu $3d_{x^2-y^2}$ hole. Using the CuO$_4$ parameters of Shimizu \textit{et al.}\ \cite{Shimizu1990},
$(1-\gamma_\infty)=21.6(4)$ and $(1-R)B_{\rm free\text{-}ion}=80(5)$~MHz, together with the fitted $^{63}$Cu quadrupole frequencies from Table~\ref{tab:ZFNMR}, we obtain
\begin{align*}
	n_d(\text{Ba}) \approx n_d(\text{Pb}) \approx n_d(\text{Sr}) = 0.20(4).
\end{align*}
Within the uncertainty of the combined errors in $\nu_Q^{\rm latt}$, $\nu_Q^{\rm exp}$, and the Sternheimer parameters, the Cu $3d$-hole occupancy is identical for all compounds. The small value of $n_d$, together with the sizable transferred hyperfine fields at $^{31}$P, confirms that $A$TCPO is a ligand-hole--dominated charge-transfer insulator in which the Cu magnetic moment is strongly delocalized onto the surrounding oxygen and phosphorus ligands.

Neutron diffraction establishes the ordered-moment size, moment direction, and the distinct interlayer stacking realized in PbTCPO versus BaTCPO/SrTCPO~\cite{Kimura2018Acation,Kimura2016_magneto}, while Cu ZFNMR provides a local constraint on the Cu-moment direction: the fitted angle $\theta$ is small in all three compounds, and for PbTCPO we obtain $\theta \simeq 0^\circ$, indicating that the Cu-site internal field is nearly aligned with the principal EFG axis. Since the point-charge EFG analysis places the $V_{zz}$ axis close to the local CuO$_4$ plaquette normal, this agrees with the neutron-derived Cu-moment direction. At the same time, the Cu-site internal fields remain comparable across BaTCPO, PbTCPO, and SrTCPO, consistent with a local Cu magnetic scale in all three compounds. In contrast, the much larger ordered-state $^{31}$P internal field in PbTCPO than in BaTCPO and SrTCPO is compatible with the ferroic stacking in Pb and the stronger interlayer cancellation in the antiferroically stacked Ba/Sr compounds. Thus, Cu ZFNMR constrains the local Cu-moment direction and magnetic scale, while $^{31}$P NMR constrains the stacking-dependent off-site field pattern, and both are consistent with the neutron diffraction results.

\vspace{128pt}

\onecolumngrid

\begin{table}[ht]
	\renewcommand{\arraystretch}{1.5}
		\caption{\label{tab:ZFNMR}%
			Observed and calculated $^{63,65}$Cu zero-field NMR transition frequencies
			in $A$TCPO at $T = 4.2$~K. ``Observed'' are experimental peak positions with
			one-standard-deviation uncertainties in parentheses. ``Calculated''
			values are obtained from the joint $^{63,65}$Cu Hamiltonian fit described in
			Sec.~\ref{subsec:CuZFNMR}.}
		\begin{ruledtabular}
			\begin{tabular}{|c|c|c c|c|c|c|}
				& & \multicolumn{2}{c|}{Frequency (MHz)} & FWHM & $\nu_L$ & $\nu_Q$ \\
				Compound : nucleus &
				Transition &
				Observed & Calculated & (MHz) & (MHz) & (MHz) \\
				\hline
				BaTCPO : $^{63}$Cu &
				$-\tfrac{3}{2}\!\leftrightarrow\!-\tfrac{1}{2}$ &
				198.355(5) & 198.38(3) & 0.61 &        &        \\
				&
				$-\tfrac{1}{2}\!\leftrightarrow\!+\tfrac{1}{2}$ &
				166.686(3) & 166.77(2) & 0.62 & 166.58(2) & 32.46(2) \\
				&
				$+\tfrac{1}{2}\!\leftrightarrow\!+\tfrac{3}{2}$ &
				135.045(4) & 134.99(3) & 0.61 &        &        \\
				& & & & & & \\
				BaTCPO : $^{65}$Cu &
				$-\tfrac{3}{2}\!\leftrightarrow\!-\tfrac{1}{2}$ &
				207.857(8) & 207.84(17) & 0.56 &        &        \\
				&
				$-\tfrac{1}{2}\!\leftrightarrow\!+\tfrac{1}{2}$ &
				178.569(5) & 178.60(12) & 0.67 & 178.45(2) & 29.99(2) \\
				&
				$+\tfrac{1}{2}\!\leftrightarrow\!+\tfrac{3}{2}$ &
				149.278(9) & 149.23(17) & 0.59 &        &        \\
				\hline
				PbTCPO : $^{63}$Cu &
				$-\tfrac{3}{2}\!\leftrightarrow\!-\tfrac{1}{2}$ &
				196.358(16) & 196.37(8) & 1.96 &        &        \\
				&
				$-\tfrac{1}{2}\!\leftrightarrow\!+\tfrac{1}{2}$ &
				163.694(8)  & 163.76(5) & 1.71 & 163.65(5) & 32.72(5) \\
				&
				$+\tfrac{1}{2}\!\leftrightarrow\!+\tfrac{3}{2}$ &
				130.870(12) & 130.94(8) & 1.79 &        &        \\
				& & & & & & \\
				PbTCPO : $^{65}$Cu &
				$-\tfrac{3}{2}\!\leftrightarrow\!-\tfrac{1}{2}$ &
				205.534(24) & 205.55(8) & 1.71 &        &        \\
				&
				$-\tfrac{1}{2}\!\leftrightarrow\!+\tfrac{1}{2}$ &
				175.485(12) & 175.40(6) & 1.62 & 175.31(6) & 30.24(5) \\
				&
				$+\tfrac{1}{2}\!\leftrightarrow\!+\tfrac{3}{2}$ &
				145.126(25) & 145.08(8) & 1.62 &        &        \\
				\hline
				SrTCPO : $^{63}$Cu &
				$-\tfrac{3}{2}\!\leftrightarrow\!-\tfrac{1}{2}$ &
				200.834(13) & 200.87(6) & 1.55 &        &        \\
				&
				$-\tfrac{1}{2}\!\leftrightarrow\!+\tfrac{1}{2}$ &
				168.041(7)  & 168.10(4) & 1.47 & 168.11(4) & 32.82(4) \\
				&
				$+\tfrac{1}{2}\!\leftrightarrow\!+\tfrac{3}{2}$ &
				135.361(10) & 135.37(6) & 1.55 &        &        \\
				& & & & & & \\
				SrTCPO : $^{65}$Cu &
				$-\tfrac{3}{2}\!\leftrightarrow\!-\tfrac{1}{2}$ &
				210.394(22) & 210.37(5) & 1.55 &        &        \\
				&
				$-\tfrac{1}{2}\!\leftrightarrow\!+\tfrac{1}{2}$ &
				180.134(10) & 180.08(4) & 1.31 & 180.08(4) & 30.32(4) \\
				&
				$+\tfrac{1}{2}\!\leftrightarrow\!+\tfrac{3}{2}$ &
				149.825(20) & 149.82(5) & 1.31 &        &        \\
			\end{tabular}
		\end{ruledtabular}
	\renewcommand{\arraystretch}{1}
\end{table}

\twocolumngrid

\section{Conclusion}

The square-cupola cuprates $A$(TiO)Cu$_4$(PO$_4$)$_4$ provide a canonical platform for cluster-multipole physics in an antiferromagnet, where the spin texture on each Cu$_4$O$_{12}$ unit is described in terms of a magnetic quadrupole. $A$-site stereochemistry controls the symmetry of interlayer coupling and thereby selects whether the quadrupole pattern stacks antiferroically (Ba/Sr) or ferroically (Pb) along the $c$ axis. In PbTCPO, ferroic stacking yields a non-canceling macroscopic multipole state that forms the basis for the linear magnetoelectric and nonreciprocal optical responses reported previously, while crystal chirality enables an additional octupolar component relevant to magnetic-field-only domain control.

Here we report an NMR characterization of PbTCPO based on single-crystal $^{31}$P NMR and zero-field $^{63,65}$Cu NMR. Single-crystal $^{31}$P NMR determines the transferred hyperfine tensor at the P sites, including its anisotropy extracted from rotation experiments. Zero-field $^{63,65}$Cu NMR yields a large internal field $B_{\rm int}=14.50(6)$~T and a quadrupole frequency $\nu_Q=32.72(5)$~MHz, fixing the magnetic and EFG environment at the Cu sites in the ordered state. Point-charge EFG calculations with Sternheimer corrections further indicate a small on-site Cu $3d$-hole occupancy $n_d=0.20(4)$, consistent with a ligand-hole--dominated charge-transfer character.

These parameters separate cleanly into observables governed by local transfer and those governed by interlayer stacking. The $^{31}$P hyperfine couplings serve as a quantitative proxy for local Cu--O--P covalency (transferred spin density), whereas the scale of the internal field at $^{31}$P is far more sensitive to the interlayer stacking of the quadrupole pattern through cancellation (or lack thereof) of contributions from neighboring layers. This offers a concrete basis for interpreting why PbTCPO exhibits enhanced $^{31}$P internal fields even when the local Cu--O--P transfer strength is not anomalously large. Together with the Cu-site $B_{\rm int}$ and EFG parameters, these results supply benchmark inputs for model Hamiltonians and for quantitative calculations of local fields and spectra in square-cupola cuprates.
More broadly, the ability to separate stacking-driven (interlayer cancellation) and covalency-driven (transferred spin density) contributions to internal fields offers a general blueprint for linking symmetry-based multipole concepts to quantitative materials modeling in correlated oxides, supporting progress in adjacent directions such as antiferromagnetic spintronics, magnonics/THz dynamics, and nonreciprocal photonics.

\section*{Acknowledgments}
We thank Dr. Eun Sang Choi for assistance with susceptibility measurements using the SQUID magnetometer. RS is thankful to the Fulbright Foundation for enabling his sabbatical at NHMFL.

The Estonian Science Council (ETAg) Grants IUT23-7, PRG4, PRG1702, and TemTA-25 supported this research.  This work was partially supported by Japan Society for the Promotion of Science (JSPS) KAKENHI (Grants-in-Aid for Scientific Research) Grants JP19H01832, JP21H01035, JP23KJ0502, JP24K00575, JP24H01599, JP25H00600, and JP25H00392, as well as by the Murata Science and Education Foundation.  

The authors acknowledge support from the ISABEL project, funded by the European Union under the Horizon 2020 Grant Agreement No. 871106. The work at NICPB was conducted using research infrastructures EstMagLab (TARISTU24-16) and AKKI (TARISTU24-15) funded by the Estonian
Research Council, and supported by the European Research Council (ERC) under the European Union Horizon 2020 research and innovation program, Grant Agreement No. 885413.
\medskip

\nocite{Rasta2026_data}
\bibliography{references}

\end{document}